\documentclass[aps,prx,reprint,superscriptaddress,longbibliography,amsmath,amssymb,notitlepage,pra]{revtex4-1}
\usepackage[utf8]{inputenc}
\usepackage{amsfonts}
\usepackage{physics}
\usepackage{graphicx}
\usepackage{overpic}
\usepackage{color}
\usepackage[dvipsnames]{xcolor}
\usepackage[hidelinks]{hyperref}
\usepackage{comment}
\usepackage{braket}

\usepackage[T1]{fontenc}
\usepackage{enumitem}
\usepackage{booktabs}
\usepackage{float}
\usepackage{dsfont}

\usepackage{soul}

\usepackage{orcidlink}
\newcommand{\orcidnora}{\orcidlink{0000-0002-9490-2536}}
\newcommand{\orciddaniel}{\orcidlink{0000-0001-7658-3546}}
\newcommand{\orciddarvin}{\orcidlink{0000-0001-8805-3761}}
\newcommand{\orcidpietro}{\orcidlink{0000-0001-5279-7064}}
\newcommand{\orcidsimone}{\orcidlink{0000-0002-8882-2169}}

\newcommand{\uulm}{Institute for Complex Quantum Systems \& Center for Integrated Quantum Science and Technology, Ulm University, Albert-Einstein-Allee 11, 89069 Ulm, Germany}
\newcommand{\unipd}{Dipartimento di Fisica e Astronomia "G. Galilei" \& Padua Quantum Technologies Research Center, Universit{\`a} degli Studi di Padova, Italy I-35131, Padova, Italy}

\newcommand{\pdinfn}{INFN, Sezione di Padova, via Marzolo 8, I-35131, Padova, Italy}

\begin{document}

\title{Finite-temperature Rydberg arrays: quantum phases and entanglement characterization}

\author{Nora Reini\'c\orcidnora}
\affiliation{\unipd}
\affiliation{\pdinfn}
\author{Daniel Jaschke\orciddaniel}
\affiliation{\unipd}
\affiliation{\pdinfn}
\affiliation{\uulm}
\author{Darvin Wanisch\orciddarvin}
\affiliation{\unipd}
\affiliation{\pdinfn}
\author{Pietro Silvi\orcidpietro}
\affiliation{\unipd}
\affiliation{\pdinfn}
\author{Simone Montangero\orcidsimone}
\affiliation{\unipd}
\affiliation{\pdinfn}
\affiliation{\uulm}

\begin{abstract}

As one of the most prominent platforms for analog quantum simulators, Rydberg atom arrays are a promising tool for exploring quantum phases and transitions. While the ground state properties of one-dimensional Rydberg systems are already thoroughly examined, we extend the analysis towards the finite-temperature scenario. For this purpose, we develop a tensor network-based numerical toolbox for constructing the quantum many-body states at thermal equilibrium, which we exploit to probe classical correlations as well as entanglement monotones. We clearly observe ordered phases continuously shrinking due to thermal fluctuations at finite system sizes. Moreover, by examining the entanglement of formation and entanglement negativity of a half-system bipartition, we numerically confirm that a conformal scaling law of entanglement extends from the zero-temperature critical points into the low-temperature regime.

\end{abstract}

\maketitle


With one or more electrons in a highly excited energy state, Rydberg atoms display strong interparticle interactions \cite{Urban2009, Low2012, Weber2017}, making them excellent candidates for programmable quantum devices \cite{Wu2021, Browaeys2020}. With the possibility of arranging individual Rydberg atoms in arrays of arbitrary geometries using optical tweezers \cite{Kaufman2021, Barredo2018, Endres2016, Manetsch2024, Pause2024, Schlosser2023}, recent advancements in quantum computing with neutral atoms have demonstrated high-fidelity qubit gates \cite{Levine2019, Fu2022, Evered2023} and arbitrary connectivity \cite{Bluvstein2022, Bluvstein2023}. Moreover, the progress in analog quantum simulation enables the preparation and control of hundreds of neutral atom qubits, allowing to probe quantum many-body properties directly through measurements \cite{Gross2017, Browaeys2020, Wurtz2023}. Quantum simulators have demonstrated to be successful in exploring the various quantum phases, critical behavior, and underlying universal properties \cite{Bernien2017, Keesling2019, Leseleuc2019, Scholl2021, Ebadi2021, Zhang2024}. 

A crucial resource for quantum technologies is entanglement, which in turn makes characterizing and quantifying the entanglement a fundamental task~\cite{Amico2008}. From the experimental perspective, even though there exist different methods for entanglement certification~\cite{Baccari2017, Friis2018}, promising approaches offered by two-copy schemes~\cite{Daley2012, Islam2015, Bluvstein2022}, or randomized measurement sampling protocols~\cite{Elben2023, Notarnicola2023, Elben2020}, measuring the entanglement still poses significant challenges. Thus, theoretical analyses and numerical simulations remain pivotal in quantifying the entanglement of quantum many-body systems~\cite{Eisert2010, Laflorencie2016}.

When considering bipartite entanglement, an important point to emphasize is the difference between pure-state and mixed-state entanglement characterization. For pure states, entanglement between the subsystems is well-defined, for example, as the entropy of the reduced density matrix of either subsystem, whether in terms of the von Neumann entropy~\cite{Bennet1996} or the Renyi entropies~\cite{Cui2013}. However, physical systems are never fully isolated from their surroundings and experimental platforms are never cooled down to absolute zero temperature, requiring to switch towards the mixed-state framework. There is no unique way of characterizing mixed-state entanglement, and different measures, each with their own different physical interpretations, have been constructed \cite{Brus2002, Plenio2007}. From a technical standpoint, estimating these measures is hard because it requires dedicated strategies, capable of discriminating quantum correlations from classical ones. Additionally, some mixed-state entanglement measures, such as entanglement of formation~\cite{Bennet1996eof}, require performing minimizations over all possible pure-state decompositions of the density matrix \cite{TTOpaper}.

Here, we numerically analyze the finite-temperature physics of a Rydberg atom chain at thermal equilibrium and characterize its bipartite entanglement properties. Towards this goal, we employ the Tree Tensor Operator (TTO) \cite{TTOpaper}, a tensor network ansatz for representing a density matrix, highly suitable for the estimation of the mixed state properties. The TTO provides a global purification of the many-body state and efficiently encodes entanglement content. In previous work \cite{TTOpaper}, thermal TTOs were constructed by manually stacking individually computed Hamiltonian eigenstates, an approach posing strict limitations on the accessible temperature range. In contrast, our algorithm performs the numerical conversion from a Locally Purified Tensor Network (LPTN)~\cite{Werner2016} to the TTO, where the LPTN is suitable for imaginary time evolution \cite{Schollwock2011}. As a result, the number of included eigenstates is simply controlled by setting the corresponding bond dimension. We estimate the system's purity and track the resilience of the system's ordered phases while increasing the temperature, and probe the behavior of entanglement negativity \cite{Vidal2002, Wichterich2010, Calabrese2014} and entanglement of formation as the mixed-state entanglement measures. Finally, we test for finite-temperature extensions of zero-temperature quantum critical scalings in the vicinity of quantum phase transitions.

The paper is structured as follows: in Section \ref{sec:method}, we explain the numerical method. Then in Section \ref{sec:ryd-hamiltonian}, we introduce the Rydberg Hamiltonian that we simulate, and give a short overview of the previous findings regarding the structure of the ground state phase diagram. We show our results in Sections \ref{sec:finite-t-ph-diag} and \ref{sec:entanglement}. Finally, we summarize the conclusions in Section \ref{sec:conclusions}. Details on the numerical method can be found in Appendices \ref{sec:lptn-tto-conversion}, \ref{sec:eof-minimization}, and \ref{sec:convergence}. 


\section{Method} \label{sec:method}

In order to construct an efficient numerical algorithm for studying finite-temperature physics, it is important to choose an appropriate representation of the density matrix.
Since the TTO representation is well-suited for the extraction of the otherwise hardly accessible mixed-state properties in a low-temperature regime, here we develop the algorithm for constructing the TTO density matrix for quantum many-body systems at thermal equilibrium.
In this section, we first provide a recap of TTOs and their advantages in representing and compressing the density matrix. Then, we provide the definitions of two mixed state entanglement measures and clarify how to extract them from a TTO. Finally, we present the main steps of our numerical algorithm.

\vspace{-0.1cm}
\subsection{Tree Tensor Operator}
\begin{figure}[t]
  \centering
    \includegraphics[width=0.47\textwidth]{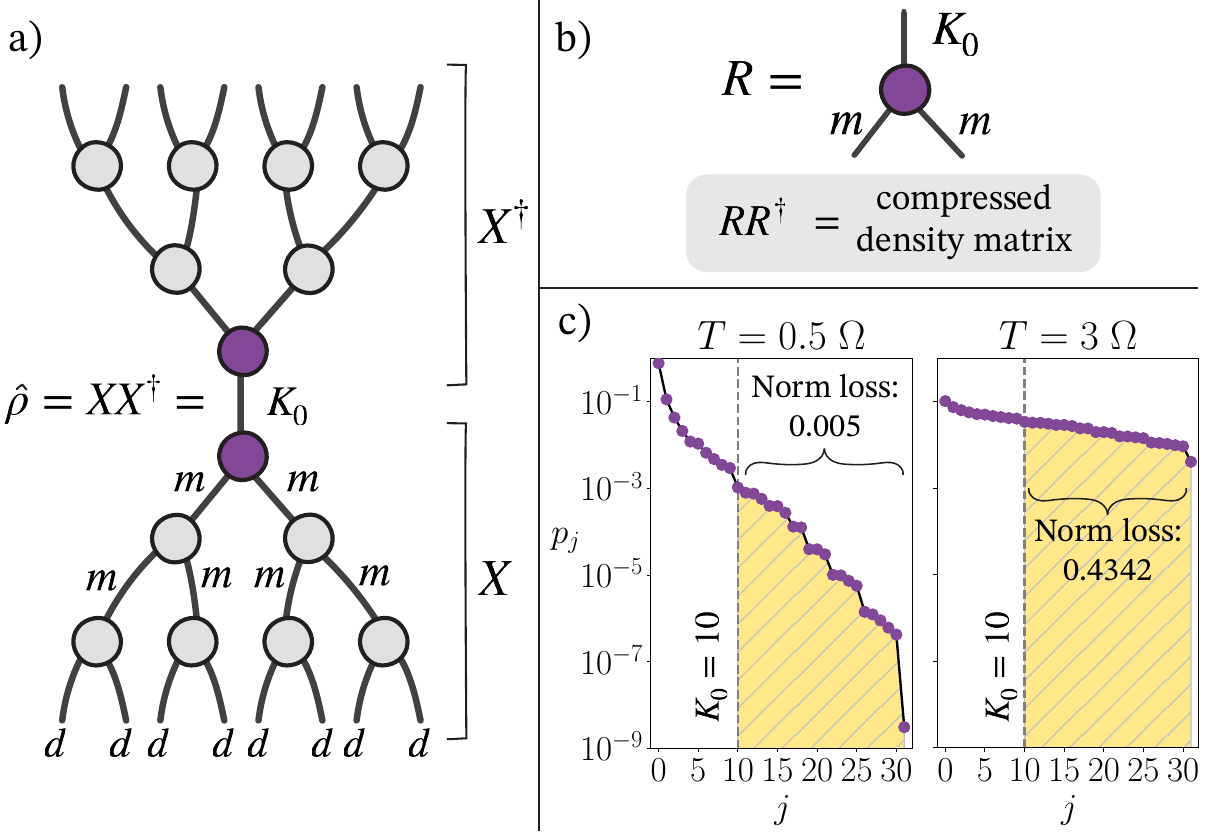} 
  \caption{\textbf{Tree Tensor Operator (TTO) density matrix}. (a) A density matrix can be decomposed as $\hat{\rho} = XX^{\dag}$, with columns of $X$ corresponding to $\sqrt{p_j} \ket{\psi_j}$. In the TTO ansatz, $X$ and $X^{\dag}$ are represented with the tree tensor network-like structures where the connecting link corresponds to the sum over the pure states $\ket{\psi_j}$. The bond dimension $m$ corresponds to the bound for system-to-system correlation, and the bond dimension $K_0$ is the number of pure states we keep in the density matrix.
  (b) When isometrized towards the uppermost tensor $R$, the remaining (gray) tensors in a TTO are local unitary transformations. Thus, all the information about bipartite entanglement between the system's halves is concentrated in the tensor $R$ of controllable dimensions, and its singular values are the square roots of mixed state probabilities $\sqrt{p_j}$.
  (c) Example comparison between thermal probabilities distribution for different temperatures $T$ and the corresponding norm loss after truncating to $K_0=10$ lowest-energy states. The norm loss is larger for the higher temperature (right) with respect to the lower temperature (left). The distributions are obtained for a chain of $N=5$ Rydberg atoms ($R_B/a=1.14, \Delta/\Omega=2.9$, see Hamiltonian in Eq.~\eqref{eq:ryd-ham}).
  }
  \label{fig:tto}
  \end{figure}

A TTO (Fig. \ref{fig:tto}a) is a natural tensor network ansatz for encoding a mixed state density matrix; by construction, it is loopless, guarantees positivity of the corresponding density matrix and, by performing isometrisation \cite{Silvi2019}, allows us to compress all information about the mixed state probabilities and bipartite entanglement into a single tensor (Fig. \ref{fig:tto}b).

The main mechanism with which TTO algorithms gain their efficiency is based on the following reasoning. A quantum system at thermal equilibrium is described with the density matrix
\begin{equation}
\begin{split}
    &\hat{\rho} = \sum_{j}p_j \ket{\psi_j}\bra{\psi_j},\ \ 
    p_j = \frac{e^{-\beta E_j}}{Z}, \\
\end{split}
\label{dm_finiteT}
\end{equation}

\noindent where $\ket{\psi_j}$ is the $j^{\mathrm{th}}$ eigenstate of the system's Hamiltonian with corresponding eigenenergy $E_j$, $p_j$ are the classical probabilities of finding a system in a state $\ket{\psi_j}$, $Z$ is the partition function, and $\beta=1/T$ with $T$ being temperature (in Boltzmann's constant units, i.e.,, $k_B=1$). Even though, in principle, the number of eigenstates that contribute to the density matrix grows exponentially with the system size, the probabilities $p_j$ decay exponentially with the eigenstate's energy. The temperature factor in the probabilities regulates the flatness of the distribution and one can see that only the low-energy states have a non-negligible contribution to the density matrix in the low-temperature regime (Fig.~\ref{fig:tto}c). In the TTO structure, the middle connecting link corresponds to the sum over eigenstates in Eq.~\eqref{dm_finiteT}, and thus we have direct access to the number of states kept in the density matrix, determined by the corresponding bond dimension $K_0$ (Fig. \ref{fig:tto}a). With the ability to keep only the first $K_0$ eigenstates and discard the information about the rest, we keep the computational cost of TTO algorithms low, while retaining the physically accurate approximation. Altogether, we employ two separate bond dimension bounds in the TTO: $m$ as the bond dimension bounding system-to-system correlations and $K_0$ as the bound for the mixing bond dimension, i.e., for system-environment correlations. 

\subsection{Extracting the mixed-state entanglement}

As described in Fig. \ref{fig:tto}b, the complete information about the bipartite entanglement between the system's halves is contained in the TTO's root tensor $R$. Thus, $RR^{\dag}$ represents a compressed density matrix from which we can compute the mixed-state entanglement measures.

\begin{figure}[t]
  \centering
    \includegraphics[width=0.4\textwidth]{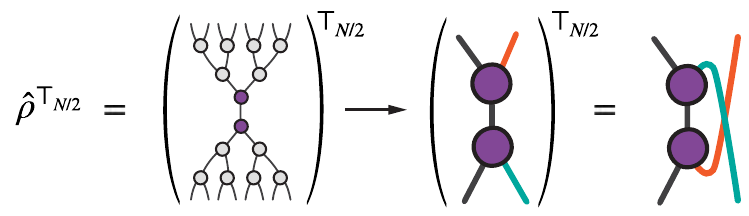}
  \caption{\textbf{Partial transpose of a TTO density matrix for computing the entanglement negativity.} Only the root tensors (purple) are relevant for extracting the entanglement negativity, thus the rest of the TTO can be discarded. Then, the partial transpose with respect to half of the system equals to flipping the corresponding legs (orange and green).}
  \label{fig:partial-transpose}
\end{figure}

Here, we compute two measures: the entanglement negativity and the entanglement of formation. The entanglement negativity is defined as

\begin{equation}
\begin{split}
    &E_N = \frac{\lVert \hat{\rho}^{\top_{N/2}} \rVert_1-1}{2},
\end{split}
\label{eq:negativity}
\end{equation}

\noindent where $\hat{\rho}^{\top_{N/2}}$ is the partial transpose of $\hat{\rho}$ with respect to half of the system and $\lVert A \rVert_1$ is the trace norm, $\lVert A \rVert_1 = Tr(\sqrt{A^{\dag}A})$. The computation of the partial transpose $\hat{\rho}^{\top_{N/2}}$ from a TTO is graphically shown in Fig.~\ref{fig:partial-transpose}.

The entanglement of formation is the mixed-state generalization of entanglement entropy. Mathematically, it is defined as a weighted sum of entanglement entropies minimized over all the possible pure-state decompositions of $\hat{\rho}$:

\begin{equation}
\begin{split}
    &E_F = \inf_{\psi_j, p_j}\Bigl\{\sum_j p_j \mathcal{S}_E (\ket{\psi_j})\ \ :\ \ \hat{\rho} = \sum_j p_j \ket{\psi_j}\bra{\psi_j}\Bigr\}, \\
\end{split}
\label{eq:entanglement-of-formation}
\end{equation}

\noindent with $\mathcal{S}_E$ denoting the entanglement entropy of half-system bipartition. In general, calculating the entanglement of formation is an extremely challenging task, and exact solutions are known only for small system sizes \cite{Concurrence1, Concurrence2} or very specific cases \cite{Terhal2000, Vollbrecht2001}. Here, we compute it using the procedure presented in Ref.~\cite{TTOpaper}, improved as described in Appendix \ref{sec:eof-minimization}.

\vspace{-0.4cm}
\subsection{Obtaining the Tree Tensor Operator density matrix}
\vspace{-0.2cm}

To access the system’s properties at an arbitrary temperature, we evolve the infinite-temperature density matrix according to the imaginary time evolution \cite{Verstraete2004,Zwolak2004,Weiss2018}. This is possible due to the fact that an infinite-temperature density matrix, $\hat{\rho}_{\infty}$, is proportional to an identity matrix, allowing us to write:

\begin{equation}
\begin{split}
    \hat{\rho} (T) &= e^{-\beta(T) \hat{H}} 
    = e^{-\beta(T) \hat{H}/2} \mathds{1} e^{-\beta(T) \hat{H}/2}\\
    &\propto e^{-\beta(T) \hat{H}/2} \hat{\rho}_{\infty} e^{-\beta(T) \hat{H}/2}.
\end{split}
\label{imag_time_evo}
\end{equation}

\begin{figure*}[t]
  \centering
    \includegraphics[width=0.75\textwidth]{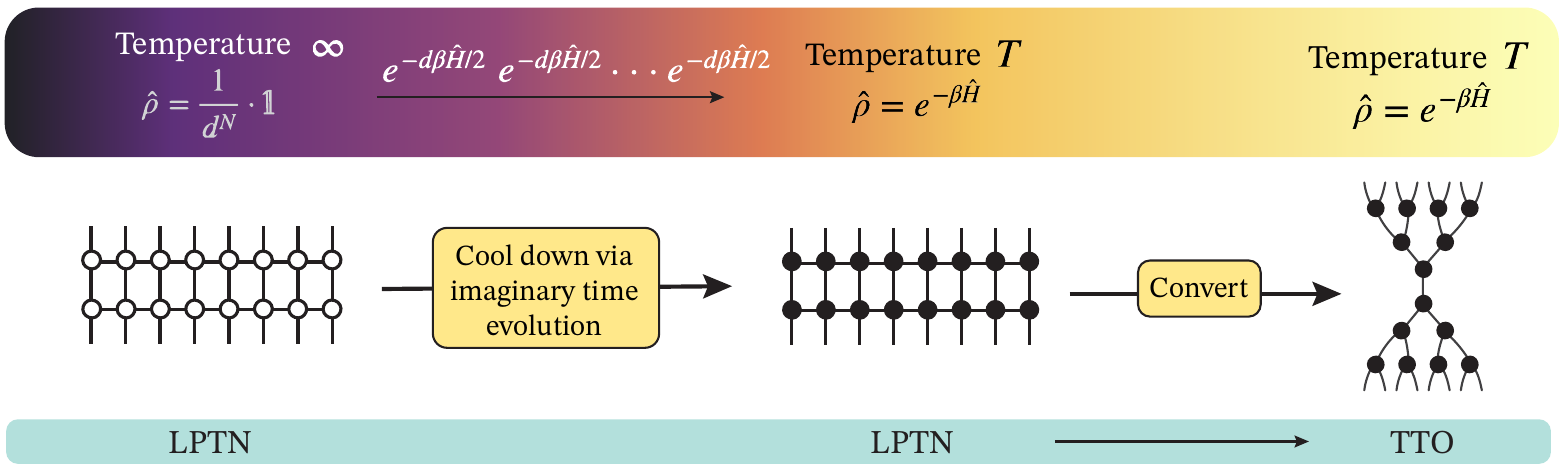}
  \caption{\textbf{An algorithm for obtaining the thermal density matrix in the TTO form}. An infinite-temperature density matrix is always proportional to an identity matrix, and it is therefore straightforward to write in the LPTN form. We take this infinite-temperature density matrix as an initial state and evolve it via the imaginary time evolution, i.e., consecutively applying the operator $e^{-\mathrm{d} \beta \hat{H}/2}$ on both sides of the LPTN. Here, $\mathrm{d}\beta/2$ is an imaginary time step which, in terms of physical interpretation, represents a small decrease in temperature. This way, we can reach a density matrix at an arbitrarily low temperature. Once we reach the desired temperature, we convert the LPTN into the TTO form with the procedure described in Appendix \ref{sec:lptn-tto-conversion}.
  }
  \label{fig:algorithm}
\end{figure*}

In the last expression from Eq. \eqref{imag_time_evo} we recognize the form of a unitary time evolution of the density matrix operator with the substitution i$t \rightarrow \beta /2$, where $t$ is a real time variable. Therefore, for obtaining the density matrix at a certain temperature we can evolve the identity matrix using the standard time evolution algorithms, with an imaginary time step.

However, we cannot perform imaginary time evolution directly on a TTO: the representation of an identity matrix as a TTO would require the $K_0$ dimension to scale exponentially with system size which prohibits the simulation in terms of computational resources. For this reason, we take a deviation via the matrix product state's mixed state counterpart - the LPTN, which encodes the identity matrix in a trivial way.

We evolve LPTN in imaginary time using the time-dependent variational principle (TDVP) \cite{Haegeman2011, Haegeman2016} algorithm. For details on the TDVP algorithm see Ref. \cite{Paeckel2019, Jaschke2018}. Once we obtain the finite-temperature LPTN density matrix, we convert it to a TTO with an iterative procedure described in Appendix \ref{sec:lptn-tto-conversion}. The whole procedure for obtaining a TTO thermal density matrix is summarized in Fig. \ref{fig:algorithm}.

The timestep we use in imaginary time evolution is $d \beta/2 =0.05$ for obtaining the finite temperature phase diagrams in Section \ref{sec:finite-t-ph-diag}. The value is chosen based on the convergence study, showing that eventual errors are the largest in the vicinity of quantum critical transitions. The value of $d\beta/2$ taken for the analysis at critical regions carried out in Section \ref{sec:entanglement} is thus further reduced to $d \beta/2 = 0.025$.  The maximal entanglement bond dimension in LPTN and TTO is $m=50$, and the maximal number of pure states we keep in the TTO density matrix is $K_0=100$. The error resulting from truncation in LPTN to TTO conversion is analyzed in Section \ref{sec:finite-t-ph-diag}.


\section{Rydberg atom Hamiltonian} \label{sec:ryd-hamiltonian}

We consider a system of two-level Rydberg atoms pinned by optical tweezers. Such system realizes a spin-$1/2$ model by associating the two spin states with the ground state $\ket{g}$ and Rydberg state $\ket{r}$ \cite{Browaeys2020}. Rydberg interactions and laser-driven excitations can be encoded through the local Pauli matrix operators $\sigma_x = \ketbra{g}{r} + \ketbra{r}{g} $ and $n_z = \ketbra{r}{r}$, and in the rotating wave approximation the resulting $N$-site Hamiltonian is:
\begin{align}
\hat{H} &= \frac{\Omega}{2} \sum_{i=1}^{N} \sigma^i_x 
            - \Delta \sum_{i=1}^{N} n^i_z 
            + \sum_{i \neq j} \frac{C_6}{R_{ij}^6} n^i_z n^j_z,
\label{eq:ryd-ham}
\end{align}

\noindent with indices $i$ and $j$ going over the system sites, $C_6$ as the Van der Waals interaction strength, and $R_{ij}$ as the distance between two sites in the lattice. The Rabi frequency $\Omega$ and detuning $\Delta$ are the laser-atom coupling parameters. The units are chosen such that the reduced Planck constant is $\hbar = 1$. It is convenient to express the long-range Rydberg interactions in terms of the Rydberg blockade radius per lattice spacing $R_B/a$, where $R_B = \sqrt[6]{ \frac{C_6}{\Omega}}$. The blockade radius $R_B$ separates a regime in which the interaction is much larger than the Rabi frequency, so the state with more than one excited atom within that radius is shifted far from resonance and thus dynamically decoupled \cite{Urban2009}. 

We study a chain of $N$ Rydberg atoms described by the Hamiltonian in Eq. \eqref{eq:ryd-ham} with open boundary conditions. In numerical simulations, we set to zero all the interactions that are more than four lattice spacings apart. We choose the lattice spacing $a$ as the unit of length and Rabi frequency $\Omega$ as the unit of energy.

As thoroughly analyzed in the literature, ground states of many-body Rydberg chains are known to exhibit a rich phase diagram \cite{Weimer2010, Bernien2017, Samajdar2018, Rader2019, Yu2022, Maceira2022}. In the regime of small detuning $\Delta/\Omega$, the leading contribution comes from the Rabi term, resulting in a low occupancy of the Rydberg state throughout a system and a disordered phase without any excitation pattern. Increasing the detuning $\Delta/ \Omega$, the system starts to favor excitations of the atoms, however, the Rydberg interaction contribution tends to suppress nearby excitations. The interplay between these two contributions results in different excitation patterns, depending on the blockade radius $R_B/a$. The emergent phases turn out to be the crystalline phases of $\mathds{Z}_p$ symmetry-breaking order (period-$p$), starting from $\mathds{Z}_2$ in the low $R_B/a$ region, and going towards the higher orders as $R_B/a$ increases \cite{Bernien2017, Yu2022}. Accordingly, the transitions between these phases belong to variety of different criticality classes. Moreover, it was shown that the Rydberg phase diagram hosts as well the incommensurate floating phases between disordered and crystalline phases \cite{Weimer2010, Rader2019, Yu2022, Maceira2022, Zhang2024}, located in the regions from above the $\mathds{Z}_3$ lobe and higher.

\begin{figure}[t]
    \includegraphics[width=0.4\textwidth]{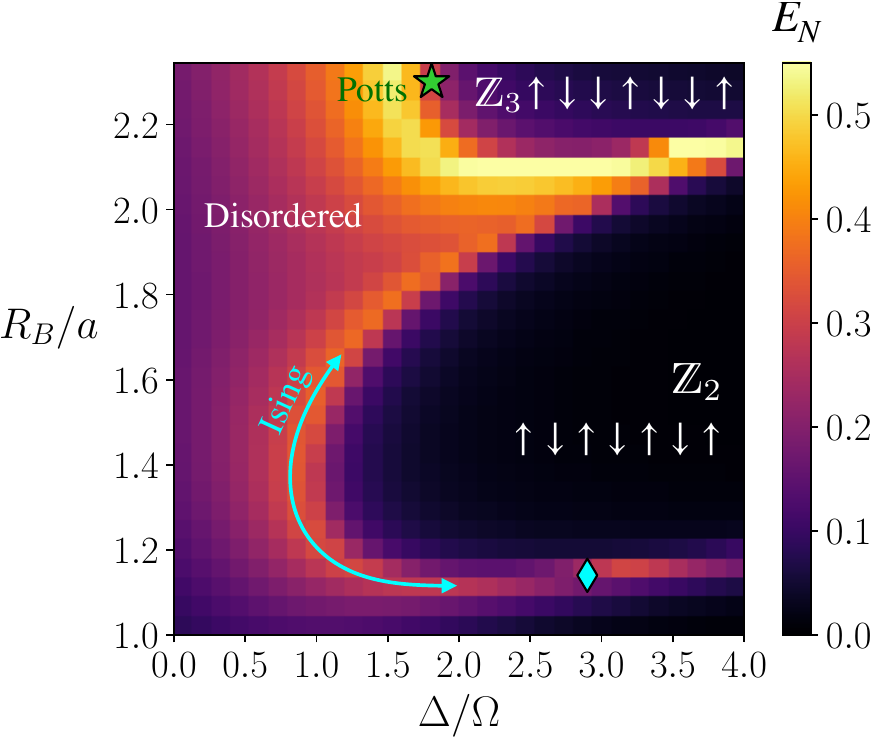} 
  \caption{\textbf{Ground state phase diagram of Rydberg atom Hamiltonian}. To detect the phase transition lines of a Rydberg atom Hamiltonian in Eq.~\eqref{eq:ryd-ham}, we measured the entanglement negativity as the function of the Rydberg blockade radius $R_B$ and detuning $\Delta$ for $N=31$ sites at temperature $T=0.025\ \Omega$. The temperature is low enough for the system to approximately be in the ground state (see Section \ref{sec:finite-t-ph-diag}). We can distinguish two crystalline phases, the ones with $\mathds{Z}_2$ and $\mathds{Z}_3$ symmetry-breaking order regimes respectively. The transition between the $\mathds{Z}_2$ and disordered phase corresponds to the Ising universality class. The green star and blue diamond mark the critical transition points that are analyzed in Section \ref{sec:entanglement}, corresponding to the three-state Potts \cite{Yu2022} and Ising universality classes, respectively. Since the position of the marked quantum critical points corresponds to its value in the thermodynamical limit, the line of maximal entanglement is slightly shifted with respect to the Potts critical point position as a consequence of the finite-size effect.}
  \label{fig:negativity-QPD}
\end{figure}

To identify the ground state phases in the parameter range of our simulations, we measure the entanglement negativity $E_N$ between two halves of the system at temperature $T=0.025\ \Omega$ in Fig.~\ref{fig:negativity-QPD}. As shown in the analysis in Section \ref{sec:finite-t-ph-diag}, this temperature is low enough for the system to be approximately in the ground state.

\begin{figure*}[t]
  \centering
    \includegraphics[width=0.9\textwidth]{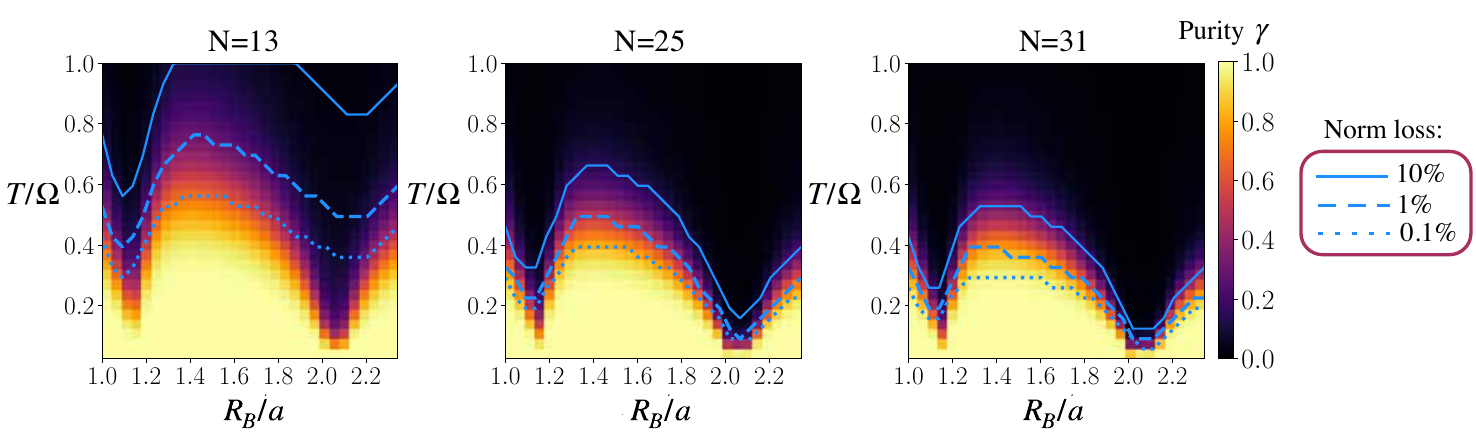} 
  \caption{\textbf{Purity computed for Rydberg atom Hamiltonian}. The behavior of the purity with temperature $T$ as the function of Rydberg blockade radius $R_B$, computed at $\Delta/\Omega = 3.034$, for $N=13$, $25$, $31$ sites. We can observe how the system transitions from the pure ground state at low temperature into the finite-temperature mixture of states. The drops in purity for certain values of $R_b/a$ correspond to the points of quantum phase transitions. Moreover, we can clearly see that as the system size increases, the ground state resilience to temperature decreases. The reason why at quantum critical transition points we can observe almost pure state even at finite temperatures is because the system size is finite. Blue curves track the accuracy of the results and denote a temperature above which the norm loss caused by truncation error becomes larger than the certain value indicated in the legend.
  }
  \label{fig:purity-cut}
\end{figure*}

Even though entanglement negativity is constructed as a mixed-state entanglement measure, here we show that it can also be used for quantifying the pure state entanglement, as the phase diagram in Fig. \ref{fig:negativity-QPD} is in line with the existing results obtained via the entanglement entropy \cite{Rader2019, Yu2022, Maceira2022}. The phase transition lines correspond to maxima in Fig. \ref{fig:negativity-QPD} where one can differentiate three phases in our parameter range: a disordered phase and two crystalline phases. It is known that the transition line between the $\mathds{Z}_2$ and disordered phase exhibits the Ising universality behavior. The transition between the $\mathds{Z}_3$ and the disordered phase is more complex, and in the parameter range of our simulations, it contains the non-conformal chiral transition line and an isolated point of the three-state Potts universality class.

To avoid edge effects and ensure the non-degeneracy of the ground state in ordered phases, in our simulations, we focus on systems of $6n + 1$ sites, which are compatible with non-degenerate configurations both for period-2 and period-3 crystalline orders. In particular, we analyze the systems of $N=13$, $25$, and $31$ sites.


\section{Finite temperature phase diagrams} \label{sec:finite-t-ph-diag}

In this section, we analyze the thermalized Rydberg systems and assess the resilience of their ground state phases to temperature. For this purpose, we compute the purity
\begin{align}
\gamma = Tr(\hat{\rho}^2) = \sum_{j=1}^{K_0} p_j^2
\label{eq:purity}
\end{align}

\noindent for every point in the phase diagram in Fig. \ref{fig:negativity-QPD} for different temperatures. Since purity can be expressed via the thermal probabilities $p_j$, it is straightforward to compute from TTO with a negligible computational cost. In the same way, with TTOs we have access to the global von Neumann entropy and Renyi entropies of any order.

The purity tells us directly about the mixedness of the system; for a pure state, the purity is equal to $1$, and the value decreases as the state becomes more mixed. Therefore, we gain a clear insight into how the ground state turns into a mixture of excited states at finite temperatures (Fig. \ref{fig:purity-cut}).

The temperature dependence of the purity for different Hamiltonian parameters reflects the structure of the low-energy part of the spectrum. In general, the most important factor deciding on the (non-degenerate) ground state's resilience to temperature is the energy gap between the ground state and the first excited state, because it directly determines the ratio of the first two largest thermal probabilities, $p_2/p_1$. Indeed, in the vicinity of quantum critical points in Fig. \ref{fig:purity-cut} where the gap decreases (and in thermodynamic limit closes), we can observe the drops in the purity. The fact that at quantum critical transition points we observe a pure state even at finite temperatures is a consequence of finite system size. As we go towards the larger system sizes, the energy levels become more dense and the transition into the finite-temperature mixed phase occurs at lower temperatures.

\begin{figure*}[t]
  \centering
    \includegraphics[width=0.7\textwidth]{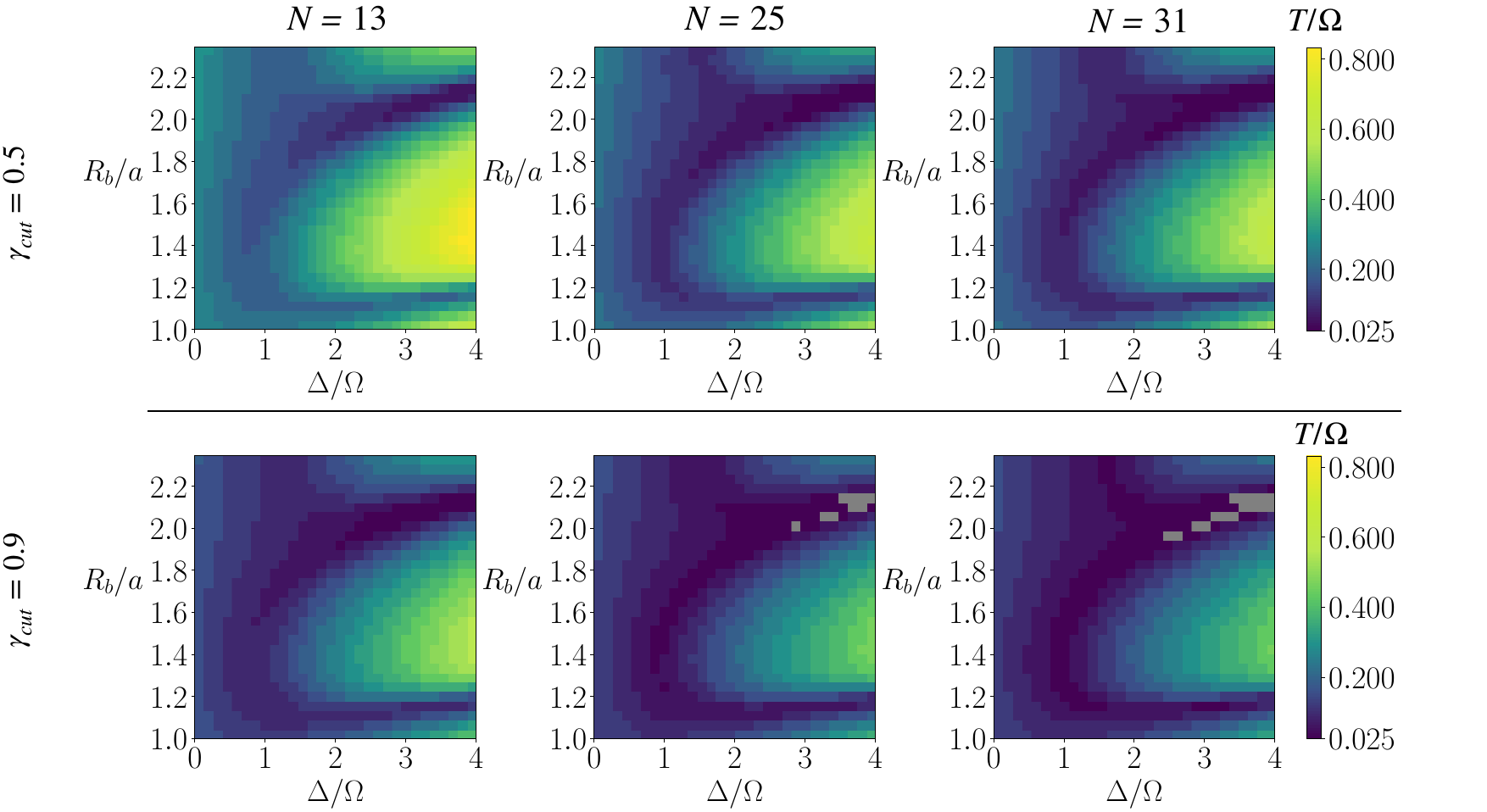} 
  \caption{
  \textbf{Finite temperature phase diagrams computed for Rydberg atom Hamiltonian}. To get an overview of how resilient the ground state ordered phases are to the temperature $T$, we plot the threshold temperature after which the purity $\gamma$ falls below the cutoff value $\gamma_{cut}=0.5$ (up) and $\gamma_{cut}=0.9$ (bottom) as a function of blockade radius $R_B/a$ and detuning $\Delta/\Omega$. The results are shown for $N=13$, $25$, $31$ sites. The temperature range is $T=[0.025,\ 1]\ \Omega$. The plots outline the features of the ground phase state diagram, and one can distinguish the $\mathds{Z}_2$ and $\mathds{Z}_3$ symmetry-breaking phases shrinking as the temperature increases. We can see that, for almost all the Hamiltonian parameter values, the temperature of $T=0.025\ \Omega$ is low enough for purity to be higher than $\gamma_{cut} = 0.9$, i.e., for $T=0.025\ \Omega$ we can expect to find the system in the ground state. The exceptions are gray regions in case of $N=25$ and $N=31$ for $\gamma_{cut}=0.9$, located between $\mathds{Z}_2$ and $\mathds{Z}_3$ lobe, for which our temperature range was not low enough to determine the threshold temperature.
  }
  \label{fig:purity-QPD}
\end{figure*}

Recall that the accuracy of our method is limited by the number of excited states we keep in a density matrix, $K_0$. This means that, when the state becomes more mixed, the state truncation error becomes larger at fixed $K_0$. An easy way of estimating the error induced by discarding the excited states is by tracking the norm loss during the LPTN to TTO conversion. The blue lines in Fig. \ref{fig:purity-cut} indicate the temperatures after which the preserved norm falls under a certain factor. As expected, the error increases with the system size for a fixed temperature value. This is because the number of excited states in the spectrum grows exponentially with the system size, and thus, increasing the system size, the number of states $K_0$ will represent a smaller ratio of all possible states. A more detailed convergence analysis is carried out in Appendix~\ref{sec:convergence}.

To get an overview of the temperature resilience of ground state phases over the entire phase diagram, we plot the threshold temperature at which the purity falls below a chosen cutoff value, across the entire Hamiltonian parameter range. The obtained finite-temperature phase diagrams for two different values of threshold purity value, $\gamma_{cut} = \{0.5, 0.9\}$, are shown in Fig. \ref{fig:purity-QPD} and one can see that they outline the features of a ground state phase diagram from Fig. \ref{fig:negativity-QPD}. The plot minima correspond to the borders between the phases, and the ground state becomes more stable as we go deeper into the crystalline phases. Again, we observe that the transition from the ground state to a mixture of excited states occurs faster for larger systems, i.e., as we go towards the thermodynamic limit.

Since analog quantum simulation experiments often target the ground state, it is instructive to compare our temperature scale to realistic temperature values in Rydberg experiments. The usual experimental Rabi frequency $\Omega$ values are within the MHz range. Taking $\Omega = 2.5$ MHz \cite{Wurtz2023}, our temperature range $[0.025$, $1]\ \Omega$ from Fig. \ref{eq:purity} translates to $[3,\ 120]\ \mu \mathrm{K}$ (or $ [0.0625,\ 2.5]$~MHz). Rydberg atom devices usually operate at $\lessapprox 10\ \mu$K ($\simeq 0.2$ MHz) \cite{Holzl2023, Wurtz2023}, and therefore our results are in line with the fact that the ground state is within the experimental reach for the majority of the Hamiltonian parameter values.
However, it should be noted that an experimental system is subject to various additional decoherence effects and motional degrees of freedom, and thus cannot necessarily be considered to be at thermal equilibrium so the temperature of an entire system is not precisely defined. Nevertheless, our results can still provide insight into the ground state resilience on a qualitative level.


\section{Mixed-state entanglement at quantum critical points} \label{sec:entanglement}

In this section, we focus on the conformal phase transition points. In particular, we compute the mixed-state entanglement monotones at two points belonging to the Ising universality class and the Potts universality class \cite{Yu2022} (marked with a blue diamond and a green star in Fig. \ref{fig:negativity-QPD}). Exploiting the TTO ansatz, we measure the temperature dependence of the bipartite entanglement between the two halves of the system via the entanglement negativity defined in Eq.~\eqref{eq:negativity} and the entanglement of formation defined in Eq.~\eqref{eq:entanglement-of-formation}. 

\begin{figure*}[t]
  \centering
    \includegraphics[width=\textwidth]{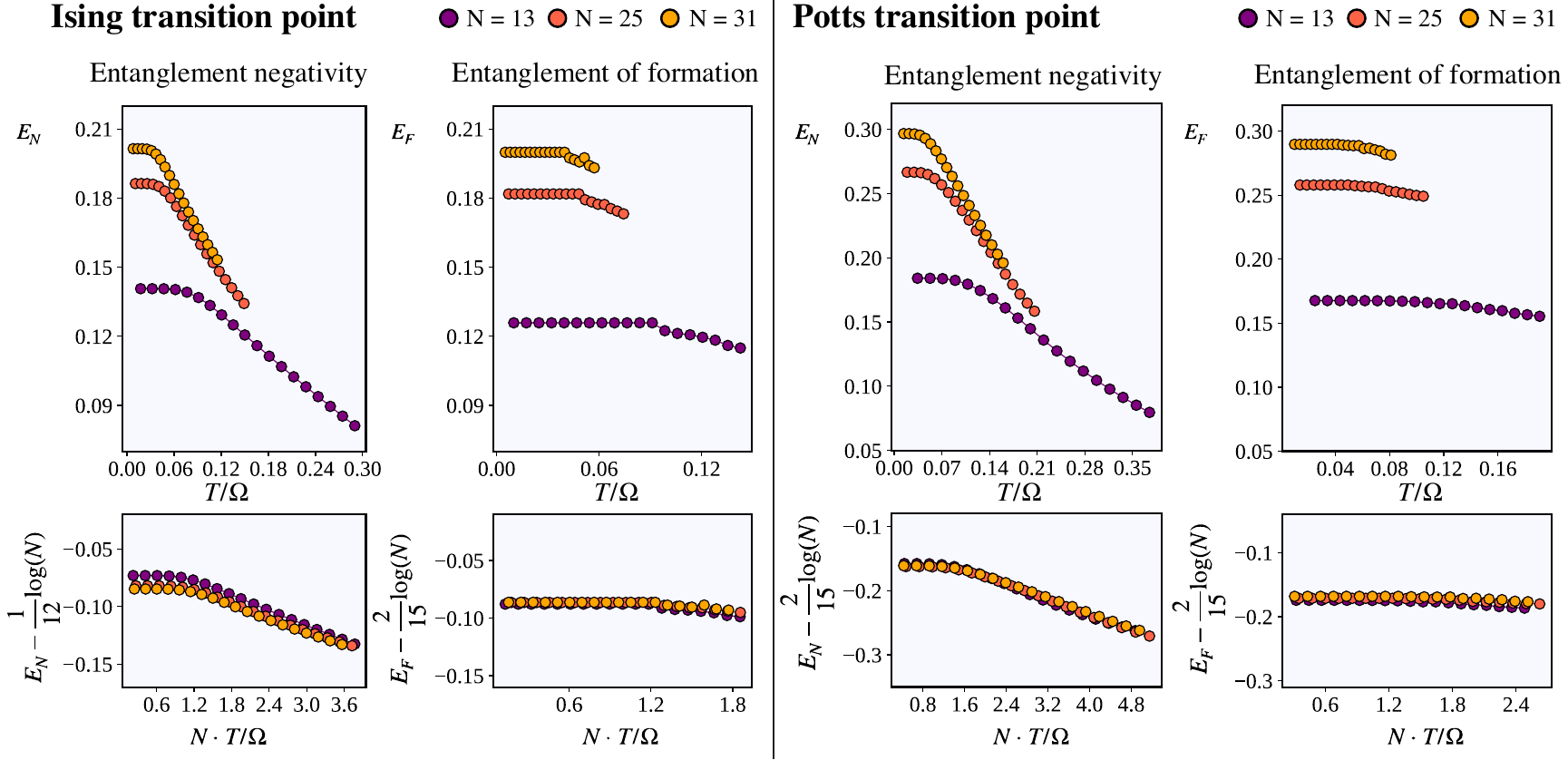} 
  \caption{
  \textbf{Thermal state entanglement at quantum critical points}. Finite-temperature decay and scaling of entanglement negativity $E_N$ and entanglement of formation $E_F$ in the vicinity of the Ising critical point at $R_B/a = 1.142$, $\Delta/\Omega=2.9$ (left) and the Potts critical point at $R_B/a = 2.3$, $\Delta/\Omega=1.808$ \cite{Yu2022} (right), computed for $N=13$, $25$, and $31$ sites. The insets show the entanglement curves after rescaling according to Eq. \eqref{eq:entanglement-scaling}. The curves show a good agreement in scaling with respect to the order of magnitude of entanglement values. The shape of the rescaled curves correspond to the functions $g_N(\cdot)$ and $g_F(\cdot)$ from Eq. \eqref{eq:entanglement-scaling}.
  }
  \label{fig:Ising-Potts-entanglement}
\end{figure*}

In order to improve the convergence of the results, we truncate the $K_0$ dimension even further and keep the first $K_0=10$ states in the density matrix in the computation of entanglement of formation. Note that, as discussed in Section \ref{sec:finite-t-ph-diag}, reducing the number of states in the density matrix increases the error in the high-temperature region. Therefore, we carry out the entanglement of formation analysis only in the low-temperature region in which the norm loss is smaller than $1\%$.

The results for Ising and Potts quantum critical points are plotted in Fig. \ref{fig:Ising-Potts-entanglement}. From both entanglement measures, we observe that entanglement is approximately constant up to a certain temperature, after which it starts to decrease. Following the analysis from Section \ref{sec:finite-t-ph-diag}, the temperature region in which the entanglement is constant corresponds to the temperature region in which the state remains approximately pure. 

The next step is to analyze and quantify finite-size effects, i.e., how entanglement scales with system size and temperature. The numerical results on the entanglement of formation $E_F$ in low-temperature Ising and Luttinger liquid XXZ model \cite{TTOpaper} predicted that in the vicinity of the conformal quantum critical point, $E_F$ should follow the logarithmic scaling. This scaling relation is the finite-temperature extension of the conformal field theory scaling for entanglement entropy. For open boundary conditions, central charge $c$, and dynamical critical exponent $z$ it reads:
\begin{equation}
\begin{split}
    &E_F(T,N) = \frac{c}{6}\mathrm{log}(N) + g_F(TN^z),
\end{split}
\label{eq:entanglement-scaling}
\end{equation}
where $g_F(\cdot)$ is a function depending on the model characteristics. Here, we neglect the corrections coming from the odd number of sites and the unequal length of the two system halves. We check if the same scaling holds in our Hamiltonian in the vicinity of the Ising and the three-state Potts transition point when the correct exponents are used ($c=1/2$, $z=1$ for Ising and $c=4/5$, $z=1$ for Potts). Moreover, we check if the analogous scaling is valid as well for entanglement negativity. Rescaling both entanglement measure curves, the results in Fig. \ref{fig:Ising-Potts-entanglement} show a good agreement with Eq. \eqref{eq:entanglement-scaling}. The slight discrepancy in the scaling of $N=13$ entanglement negativity at the Ising critical point is most likely due to the finite system size. Altogether, Eq. \eqref{eq:entanglement-scaling} represents a tool for estimating the many-body mixed-state entanglement measures in the low-temperature region of conformal critical points, based on the results for small system size.


\section{Conclusions} \label{sec:conclusions}

We have presented a tensor network algorithm for obtaining a thermal many-body density matrix and exploited our approach to study the finite-temperature effects on one-dimensional Rydberg atom systems of up to 31 particles. By computing the purity for different Hamiltonian parameters, we have studied the transition from the low-temperature pure state to the finite-temperature mixed state and demonstrated how the system's response to the temperature depends strongly on the structure of the low-energy part of the spectrum. Indeed, the transition from the gapped to the gapless point in the phase diagram is characterized by sudden drops in the property's resilience to the temperature. By plotting the temperature resilience of the system's purity over the Hamiltonian parameters, we have identified the crystalline $\mathds{Z}_2$ and $\mathds{Z}_3$ - symmetry-breaking phases and clearly observed how they shrink when increasing the temperature. We have recognized the ground state phase transition curves as the curves where the system's properties are the most sensitive to temperature. Moreover, as expected, larger systems have shown higher sensitivity to temperature effects in comparison to smaller systems.

The Rydberg chain ground state phase diagram hosts a variety of different criticality classes, and we have analyzed the temperature effects on the entanglement at the Ising and the three-state Potts quantum critical points using the mixed-state entanglement monotones negativity and entanglement of formation. Both monotones have shown the same behavior where entanglement stays unaltered up to a certain temperature and then decreases towards higher temperatures. By performing a finite-size scaling analysis of both entanglement monotones, we have provided numerical evidence that the entanglement scaling law at conformal critical points extends from the zero temperature point to the low-temperature region. Therefore, this scaling law can be used to estimate the many-body mixed-state entanglement for large system sizes, based on the results obtained for a small system size. 

Overall, we have stressed that addressing finite temperature equilibrium problems requires dedicated algorithms. From a computational perspective, simulating finite-temperature, i.e., mixed many-body quantum states, represents an additional difficulty in comparison to the already computationally demanding zero-temperature quantum many-body simulations. Our algorithm represents an efficient way of obtaining the density matrix and provides adaptability to convert from LPTN geometry, suitable for imaginary time evolution, to TTO geometry, suitable for characterization of the mixed state properties. The procedure is flexible to use beyond Rydberg systems and therefore can be employed as a toolbox for studying finite temperature physics of a general quantum many-body model. We expect that the most efficient approach to extending the framework towards two-dimensional system analysis is by mapping a model to a corresponding one-dimensional one \cite{Cataldi2021}. Another alternative is the conversion of a projected entangled-pair operator \cite{Pirvu2010, Alhambra2021}, i.e., a density matrix in projected entangled-pair state form, into a TTO.


\section*{Data and code availability} \label{sec:data-availability}

The simulations were performed using the Quantum Green Tea software version 0.3.23 and Quantum Tea Leaves version 0.4.46 \cite{QuantumTea2024}. All the simulation scripts, full datasets, and metadata are available on Zenodo \cite{Zenodo}, and all the figures are available at \cite{Figshare}.


\section*{Acknowledgments}
We thank Simone Notarnicola and Tom Manovitz for the discussion on the finite temperature effects in the experimental setup, and Lorenzo Maffi for the discussion on the conformal scaling of entanglement. The research leading to these results has received funding from the following organizations: the European Union via the Horizon 2020 research and innovation programme under the Marie Skłodowska-Curie grant agreement no. 101034319, via the NextGenerationEU project CN00000013 - Italian Research Center on HPC, Big Data and Quantum Computing (ICSC), via the H2020 projects EuRyQa and TEXTAROSSA via QuantERA2017 project QuantHEP, via QuantERA2021 project T-NiSQ, and via the Quantum Technology Flagship project PASQuanS2; the Italian Ministry of University and Research (MUR) via PRIN2022 project TANQU, and via the Departments of Excellence grant 2023-2027 Quantum Frontiers; the German Federal Ministry of Education and Research (BMBF) via the funding program quantum technologies - from basic research to market - project QRydDemo; the World Class Research Infrastructure - Quantum Computing and Simulation Center (QCSC) of Padova University. We acknowledge computational resources by Cineca on the Leonardo machine.


\appendix

\section{LPTN to TTO conversion} \label{sec:lptn-tto-conversion}

\begin{figure}[t]
  \centering
    \includegraphics[width=0.48\textwidth]{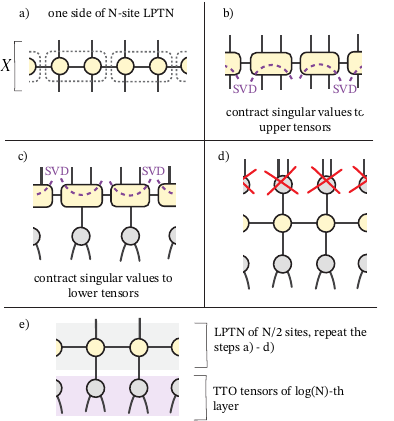} 
  \caption{
  \textbf{Conversion from Locally Purified Tensor Network to Tree Tensor Operator}.
  Black dotted rectangles denote the contraction of the enclosed tensors. The purple dashed line defines the bipartition of links for singular value decomposition (SVD). Moreover, the gray color of a tensor denotes that the tensor is unitary with respect to the appropriate bipartition of links defined by the preceding SVD. 
  }
  \label{fig:lptn-to-tto}
\end{figure}

Here, we explain the algorithm for converting the LPTN density matrix into TTO form. Since both LPTN and TTO contain two sides of the network that are Hermitian conjugated to each other, referred to as $X$ and $X^{\dag}$ in Fig. \ref{fig:tto}, it is enough to discard one of them and perform all the computations on only one of the sides. The procedure follows the steps in Fig. \ref{fig:lptn-to-tto}. Starting from the $N$-site LPTN, where $N=2^n$, $n$ being an integer, we first contract pairs of tensors (Fig. \ref{fig:lptn-to-tto}a). Then, we apply a truncated singular value decomposition (SVD) to the obtained tensors over a bipartition of links as indicated with purple dashed lines in Fig. \ref{fig:lptn-to-tto}b. The singular values are contracted to the upper tensors, and the lower tensors will remain unitary with respect to the corresponding bipartition of links. These lower tensors represent the log($N$)-th layer of the TTO. To create a base for building the next layer, we again apply a truncated SVD to the yellow tensors according to the specified bipartition (Fig. \ref{fig:lptn-to-tto}c), and this time contract the singular values to the lower tensors. In this case, the upper tensors stay unitary. Since the total density matrix contains as well the Hermitian conjugate part contracted over the upper bonds, we can ignore these upper unitary tensors because they contract with their Hermitian conjugates to an identity (Fig.~\ref{fig:lptn-to-tto}d). We obtain the tensor network form as in Fig.~\ref{fig:lptn-to-tto}e. The upper layer of tensors in Fig.~\ref{fig:lptn-to-tto}e has the LPTN-like form of half as many particles. Therefore, we can iterate the described procedure to build the rest of the TTO layers. Notice that this procedure by its construction guarantees that the isometry center is installed to the uppermost tensor. Moreover, to increase the accuracy of our method we make sure that at each singular value truncation, the isometry center of LPTN is placed at that tensor.

As noted above, TTOs are compatible only with the number of sites that are powers of two, however, the system sizes needed to realize the period-2 and period-3 phases in the Rydberg phase diagram are not powers of two. To overcome this issue, we run the imaginary time evolution on the odd-sited LPTN, and then insert the dummy sites decoupled from the rest of the system on the LPTN's edges. This way, the sites are artificially padded on both ends of the LPTN such that the new total number of sites is a power of two, without affecting any of the properties that we measure.


\section{Entanglement of formation minimization} \label{sec:eof-minimization}

To minimize entanglement of formation in Eq. \eqref{eq:entanglement-of-formation}, we need to find the optimal pure state decomposition of the density matrix. As explained in Ref.~\cite{TTOpaper}, a certain pure state decomposition of a density matrix is parametrized by a unitary matrix applied to TTO's uppermost tensor. Therefore, the optimization time depends heavily on how we sample the unitary matrices. 

Here, we improve the method proposed in Ref.~\cite{TTOpaper}, which relied on constructing the unitary matrix as an exponential of the Hermitian matrix. The drawback of such an approach is that a small change in a parameter of the Hermitian matrix can, in general, result in a large change in the elements of the corresponding unitary matrix. 

We solve this issue by using the Nelder-Mead algorithm, which is a simplex-based optimization method. We construct each point of the simplex by multiplying the initial guess for the unitary matrix with \textit{squashed} Haar-random unitary matrices, defined as

\begin{equation}
\begin{split}
    &U_{\mathrm{squash}} =U^{\dagger} D^\alpha U,
\end{split}
\label{eq:squashed-haar}
\end{equation}

\noindent where $U_{\mathrm{Haar}}=U^\dagger D U$ is an eigenvalue decomposition of the Haar-random unitary matrix \cite{Mezzadri2007}, and $\alpha \in [0,1]$. This way, we ensure that in every iteration of the search algorithm, the elements of a new guess for the unitary matrix are arbitrarily close to the elements of a previous guess, controlled with the parameter $\alpha$.

Moreover, using physical intuition we can pick a convenient initial guess for the unitary matrix based on the following reasoning. The properties of the density matrix change continuously with increasing temperature. Therefore, if we run the entanglement of formation computation for temperature points starting from the lowest temperature, we can choose the initial guess of every new point to be the optimized unitary matrix from a previous point. Such a choice additionally speeds up the computation.


\section{Convergence analysis} \label{sec:convergence}

In this section, we show how the convergence of the LPTN to TTO conversion algorithm depends on the bond dimension $m$. In the analysis, we study the system of $N=31$ sites at the Ising and the Potts critical points. We choose to showcase the largest system size and critical states because these points are the computationally most demanding task of this work. We generate LPTN states with imaginary time evolution timestep $dt=0.025$, in the temperature regime $T/\Omega = [0.02, 0.4]$. To provide a meaningful analysis of the convergence of the LPTN to TTO conversion scheme, we make sure that the LPTN state from which we start the conversion is converged in the bond dimension. By increasing the bond dimension from $m=30$ to $m=60$, values of LPTN energy remain unchanged up to 8$^{\mathrm{th}}$ digit for every temperature point. Therefore, we assume that the states of these bond dimensions are converged.

\begin{figure}[t]
  \centering
    \includegraphics[width=0.5\textwidth]{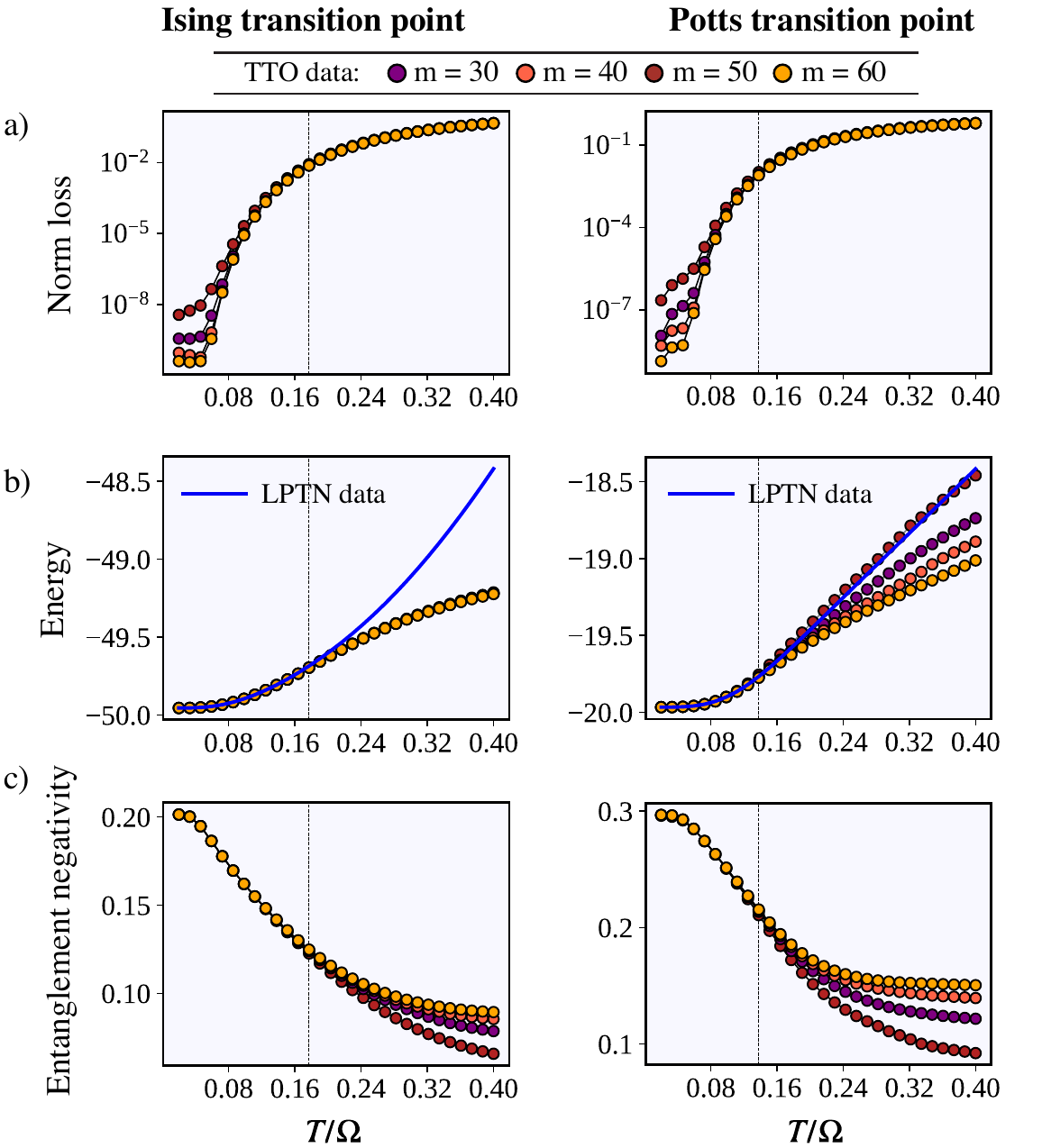} 
  \caption{
  \textbf{Convergence of LPTN to TTO conversion algorithm with the bond dimension $m$}. Norm loss during the conversion (a), energy (b), and entanglement negativity (c) are computed at the Ising and the Potts critical points of the Rydberg model for $N=31$ sites. The bond dimension $m$ of the TTO and the corresponding LPTN state are the same. The bond dimension $K_0$ is fixed to $K_0=100$. The vertical line in all plots denotes a threshold temperature below which the norm loss during the LPTN to TTO conversion is smaller than 1\%. The LPTN energy curve (blue curve) in (b) represents the reference values computed for $m=60$.}
  \label{fig:convergence}
\end{figure}

In the LPTN to TTO conversion, we fix the bond dimension $K_0=100$, i.e., we keep the first $100$ states in a density matrix. We compare the data for bond dimensions $m=[30$, $40$, $50$, $60$]. To measure the accuracy of the conversion, we track the state's norm loss due to singular value truncations during the conversion (Fig.~\ref{fig:convergence}a) and compare the energy of the resulting TTO with the corresponding energy of the LPTN (Fig.~\ref{fig:convergence}b). To give an intuition on how this reflects on the convergence of entanglement results, we show the convergence of entanglement negativity for different bond dimensions $m$ (Fig.~\ref{fig:convergence}c).

Before continuing to the convergence analysis results, it is useful to highlight the different roles of the bond dimensions $m$ and $K_0$ in a TTO. Interpretation of $K_0$ is straightforward, as $K_0$ is simply the final number of pure states we keep in the density matrix. Interpretation of $m$ is, however, much more subtle: if one thinks of a TTO as multiple pure TTN states stacked together, TTO's $m$ contains entanglement distributed over many pure states. Thus, a required $m$ depends on the amount of entanglement of each pure state and overall mixedness of a state - more entanglement in each pure state requires larger $m$, but also larger mixedness requires larger $m$ as there are more pure states to be covered. The distribution between the two contributions within $m$ is not transparent.

Back to convergence, as already discussed in Sec. \ref{sec:finite-t-ph-diag}, increasing the temperature means increasing the number of excited states that contribute to the density matrix. Therefore, the accuracy of results decreases as the number of states $K_0$ becomes insufficient to describe a physical state. This reflects directly on the norm loss in Fig.~\ref{fig:convergence}a. As we can see, the norm loss curves do not significantly vary with the bond dimension $m$, thus for these bond dimensions it is the choice of $K_0$ that sets the limit to the overall accuracy of the results. Note that the norm loss would be influenced by $m$ for smaller values of $m$. We denote the regime in which the norm loss is smaller than 1\% with a black vertical line, physically representing an approximate temperature at which $K_0=100$ states are enough to describe a thermal density matrix. Therefore, we expect our results to be accurate below this threshold temperature. Since the norm loss and the threshold temperature depend on the system size and the point in the phase diagram, we show the norm loss lines as well in Fig.~\ref{fig:purity-cut}.

By looking at the energy curves in Fig.~\ref{fig:convergence}b, the TTO results in the region below the threshold temperature coincide with the LPTN reference values. Also, the entanglement negativity in Fig.~\ref{fig:convergence}c is well-converged with $m$ within this region. The energy and the negativity at the Potts point and the negativity at the Ising point appear not to be converged with $m$ in the temperature regions above the threshold temperature. We attribute this to the above-discussed effect where larger mixedness of the system requires a larger bond dimension. Nevertheless, this region is not of interest because the accuracy is limited by $K_0$ and not $m$, therefore we do not expect accurate results there. An additional remark regarding the convergence results is that, even though the $m=30$ energy curve for the Potts point coincides with the reference LPTN curve in the plot, we expect this to be an accidental artifact of unconvergence in $m$.

In this analysis, we keep the bond dimension $m$ the same for both LPTN and TTO states, but in general, they do not have to be the same. A good choice is that $m_{\mathrm{TTO}}$ is larger or equal to $m_{\mathrm{LPTN}}$. Overall, we suggest the following strategy for picking the bond dimensions for LPTN to TTO conversion of thermal states:
\begin{enumerate}
    \item Choose $m_{\mathrm{LPTN}}$ such that LPTN states are converged.
    
    \item Pick $K_0$ and tune $m_{\mathrm{TTO}}$ such that the norm loss has converged in $m_{\mathrm{TTO}}$.

    \item Read out the threshold temperature below which the norm loss is smaller than the desired precision.

    \item If the threshold temperature is too low for the purpose of the simulation, repeat the steps $1.-3.$ with a larger $K_0$.
    
    \item Double-check if the properties have converged with $m_{\mathrm{TTO}}$ in region below the threshold temperature.

\end{enumerate}
Note that, in practice, we cannot tune the threshold temperature to drastically higher, as the required number of states $K_0$ soon becomes too large due to energies in the spectrum being dense or clustered in bands. In general, the available temperature range is of the order of magnitude of the gap between the ground and the first excited state.

\bibliography{refs}

\begin{thebibliography}{70}%
\makeatletter
\providecommand \@ifxundefined [1]{%
 \@ifx{#1\undefined}
}%
\providecommand \@ifnum [1]{%
 \ifnum #1\expandafter \@firstoftwo
 \else \expandafter \@secondoftwo
 \fi
}%
\providecommand \@ifx [1]{%
 \ifx #1\expandafter \@firstoftwo
 \else \expandafter \@secondoftwo
 \fi
}%
\providecommand \natexlab [1]{#1}%
\providecommand \enquote  [1]{``#1''}%
\providecommand \bibnamefont  [1]{#1}%
\providecommand \bibfnamefont [1]{#1}%
\providecommand \citenamefont [1]{#1}%
\providecommand \href@noop [0]{\@secondoftwo}%
\providecommand \href [0]{\begingroup \@sanitize@url \@href}%
\providecommand \@href[1]{\@@startlink{#1}\@@href}%
\providecommand \@@href[1]{\endgroup#1\@@endlink}%
\providecommand \@sanitize@url [0]{\catcode `\\12\catcode `\$12\catcode `\&12\catcode `\#12\catcode `\^12\catcode `\_12\catcode `\%12\relax}%
\providecommand \@@startlink[1]{}%
\providecommand \@@endlink[0]{}%
\providecommand \url  [0]{\begingroup\@sanitize@url \@url }%
\providecommand \@url [1]{\endgroup\@href {#1}{\urlprefix }}%
\providecommand \urlprefix  [0]{URL }%
\providecommand \Eprint [0]{\href }%
\providecommand \doibase [0]{http://dx.doi.org/}%
\providecommand \selectlanguage [0]{\@gobble}%
\providecommand \bibinfo  [0]{\@secondoftwo}%
\providecommand \bibfield  [0]{\@secondoftwo}%
\providecommand \translation [1]{[#1]}%
\providecommand \BibitemOpen [0]{}%
\providecommand \bibitemStop [0]{}%
\providecommand \bibitemNoStop [0]{.\EOS\space}%
\providecommand \EOS [0]{\spacefactor3000\relax}%
\providecommand \BibitemShut  [1]{\csname bibitem#1\endcsname}%
\let\auto@bib@innerbib\@empty
\bibitem [{\citenamefont {Urban}\ \emph {et~al.}(2009)\citenamefont {Urban}, \citenamefont {Johnson}, \citenamefont {Henage}, \citenamefont {Isenhower}, \citenamefont {Yavuz}, \citenamefont {Walker},\ and\ \citenamefont {Saffman}}]{Urban2009}%
  \BibitemOpen
  \bibfield  {author} {\bibinfo {author} {\bibfnamefont {E.}~\bibnamefont {Urban}}, \bibinfo {author} {\bibfnamefont {T.~A.}\ \bibnamefont {Johnson}}, \bibinfo {author} {\bibfnamefont {T.}~\bibnamefont {Henage}}, \bibinfo {author} {\bibfnamefont {L.}~\bibnamefont {Isenhower}}, \bibinfo {author} {\bibfnamefont {D.~D.}\ \bibnamefont {Yavuz}}, \bibinfo {author} {\bibfnamefont {T.~G.}\ \bibnamefont {Walker}}, \ and\ \bibinfo {author} {\bibfnamefont {M.}~\bibnamefont {Saffman}},\ }\bibfield  {title} {\enquote {\bibinfo {title} {Observation of {Rydberg} blockade between two atoms},}\ }\href {https://doi.org/10.1038/nphys1178} {\bibfield  {journal} {\bibinfo  {journal} {Nature Physics}\ }\textbf {\bibinfo {volume} {5}},\ \bibinfo {pages} {110–114} (\bibinfo {year} {2009})}\BibitemShut {NoStop}%
\bibitem [{\citenamefont {L\"{o}w}\ \emph {et~al.}(2012)\citenamefont {L\"{o}w}, \citenamefont {Weimer}, \citenamefont {Nipper}, \citenamefont {Balewski}, \citenamefont {Butscher}, \citenamefont {B\"{u}chler},\ and\ \citenamefont {Pfau}}]{Low2012}%
  \BibitemOpen
  \bibfield  {author} {\bibinfo {author} {\bibfnamefont {Robert}\ \bibnamefont {L\"{o}w}}, \bibinfo {author} {\bibfnamefont {Hendrik}\ \bibnamefont {Weimer}}, \bibinfo {author} {\bibfnamefont {Johannes}\ \bibnamefont {Nipper}}, \bibinfo {author} {\bibfnamefont {Jonathan~B.}\ \bibnamefont {Balewski}}, \bibinfo {author} {\bibfnamefont {Bj\"{o}rn}\ \bibnamefont {Butscher}}, \bibinfo {author} {\bibfnamefont {Hans~Peter}\ \bibnamefont {B\"{u}chler}}, \ and\ \bibinfo {author} {\bibfnamefont {Tilman}\ \bibnamefont {Pfau}},\ }\bibfield  {title} {\enquote {\bibinfo {title} {An experimental and theoretical guide to strongly interacting {Rydberg} gases},}\ }\href {https://doi.org/10.1088/0953-4075/45/11/113001} {\bibfield  {journal} {\bibinfo  {journal} {J. Phys. B: At. Mol. Opt. Phys.}\ }\textbf {\bibinfo {volume} {45}} (\bibinfo {year} {2012})}\BibitemShut {NoStop}%
\bibitem [{\citenamefont {Weber}\ \emph {et~al.}(2017)\citenamefont {Weber}, \citenamefont {Tresp}, \citenamefont {Menke}, \citenamefont {Urvoy}, \citenamefont {Firstenberg}, \citenamefont {Büchler},\ and\ \citenamefont {Hofferberth}}]{Weber2017}%
  \BibitemOpen
  \bibfield  {author} {\bibinfo {author} {\bibfnamefont {Sebastian}\ \bibnamefont {Weber}}, \bibinfo {author} {\bibfnamefont {Christoph}\ \bibnamefont {Tresp}}, \bibinfo {author} {\bibfnamefont {Henri}\ \bibnamefont {Menke}}, \bibinfo {author} {\bibfnamefont {Alban}\ \bibnamefont {Urvoy}}, \bibinfo {author} {\bibfnamefont {Ofer}\ \bibnamefont {Firstenberg}}, \bibinfo {author} {\bibfnamefont {Hans~Peter}\ \bibnamefont {Büchler}}, \ and\ \bibinfo {author} {\bibfnamefont {Sebastian}\ \bibnamefont {Hofferberth}},\ }\bibfield  {title} {\enquote {\bibinfo {title} {Calculation of {Rydberg} interaction potentials},}\ }\href {https://doi.org/10.1088/1361-6455/aa743a} {\bibfield  {journal} {\bibinfo  {journal} {J. Phys. B: At. Mol. Opt. Phys.}\ }\textbf {\bibinfo {volume} {50}} (\bibinfo {year} {2017})}\BibitemShut {NoStop}%
\bibitem [{\citenamefont {Wu}\ \emph {et~al.}(2021)\citenamefont {Wu}, \citenamefont {Liang}, \citenamefont {Tian}, \citenamefont {Yang}, \citenamefont {Chen}, \citenamefont {Liu}, \citenamefont {Tey},\ and\ \citenamefont {You}}]{Wu2021}%
  \BibitemOpen
  \bibfield  {author} {\bibinfo {author} {\bibfnamefont {Xiaoling}\ \bibnamefont {Wu}}, \bibinfo {author} {\bibfnamefont {Xinhui}\ \bibnamefont {Liang}}, \bibinfo {author} {\bibfnamefont {Yaoqi}\ \bibnamefont {Tian}}, \bibinfo {author} {\bibfnamefont {Fan}\ \bibnamefont {Yang}}, \bibinfo {author} {\bibfnamefont {Cheng}\ \bibnamefont {Chen}}, \bibinfo {author} {\bibfnamefont {Yong-Chun}\ \bibnamefont {Liu}}, \bibinfo {author} {\bibfnamefont {Meng~Khoon}\ \bibnamefont {Tey}}, \ and\ \bibinfo {author} {\bibfnamefont {Li}~\bibnamefont {You}},\ }\bibfield  {title} {\enquote {\bibinfo {title} {A concise review of {Rydberg} atom based quantum computation and quantum simulation},}\ }\href {https://doi.org/10.1088/1674-1056/abd76f} {\bibfield  {journal} {\bibinfo  {journal} {Chinese Physics B}\ }\textbf {\bibinfo {volume} {30}} (\bibinfo {year} {2021})}\BibitemShut {NoStop}%
\bibitem [{\citenamefont {Browaeys}\ and\ \citenamefont {Lahaye}(2020)}]{Browaeys2020}%
  \BibitemOpen
  \bibfield  {author} {\bibinfo {author} {\bibfnamefont {Antoine}\ \bibnamefont {Browaeys}}\ and\ \bibinfo {author} {\bibfnamefont {Thierry}\ \bibnamefont {Lahaye}},\ }\bibfield  {title} {\enquote {\bibinfo {title} {Many-body physics with individually controlled {Rydberg} atoms},}\ }\href {https://doi.org/10.1038/s41567-019-0733-z} {\bibfield  {journal} {\bibinfo  {journal} {Nature Physics}\ }\textbf {\bibinfo {volume} {16}},\ \bibinfo {pages} {132--142} (\bibinfo {year} {2020})}\BibitemShut {NoStop}%
\bibitem [{\citenamefont {Kaufman}\ and\ \citenamefont {Ni}(2021)}]{Kaufman2021}%
  \BibitemOpen
  \bibfield  {author} {\bibinfo {author} {\bibfnamefont {Adam~M.}\ \bibnamefont {Kaufman}}\ and\ \bibinfo {author} {\bibfnamefont {Kang-Kuen}\ \bibnamefont {Ni}},\ }\bibfield  {title} {\enquote {\bibinfo {title} {Quantum science with optical tweezer arrays of ultracold atoms and molecules},}\ }\href {https://doi.org/10.1038/s41567-021-01357-2} {\bibfield  {journal} {\bibinfo  {journal} {Nature Physics}\ }\textbf {\bibinfo {volume} {17}},\ \bibinfo {pages} {1324--1333} (\bibinfo {year} {2021})}\BibitemShut {NoStop}%
\bibitem [{\citenamefont {Barredo}\ \emph {et~al.}(2018)\citenamefont {Barredo}, \citenamefont {Lienhard}, \citenamefont {de~Léséleuc}, \citenamefont {Lahaye},\ and\ \citenamefont {Browaeys}}]{Barredo2018}%
  \BibitemOpen
  \bibfield  {author} {\bibinfo {author} {\bibfnamefont {Daniel}\ \bibnamefont {Barredo}}, \bibinfo {author} {\bibfnamefont {Vincent}\ \bibnamefont {Lienhard}}, \bibinfo {author} {\bibfnamefont {Sylvain}\ \bibnamefont {de~Léséleuc}}, \bibinfo {author} {\bibfnamefont {Thierry}\ \bibnamefont {Lahaye}}, \ and\ \bibinfo {author} {\bibfnamefont {Antoine}\ \bibnamefont {Browaeys}},\ }\bibfield  {title} {\enquote {\bibinfo {title} {Synthetic three-dimensional atomic structures assembled atom by atom},}\ }\href {https://doi.org/10.1038/s41586-018-0450-2} {\bibfield  {journal} {\bibinfo  {journal} {Nature}\ }\textbf {\bibinfo {volume} {561}},\ \bibinfo {pages} {79--82} (\bibinfo {year} {2018})}\BibitemShut {NoStop}%
\bibitem [{\citenamefont {Endres}\ \emph {et~al.}(2016)\citenamefont {Endres}, \citenamefont {Bernien}, \citenamefont {Keesling}, \citenamefont {Levine}, \citenamefont {Anschuetz}, \citenamefont {Krajenbrink}, \citenamefont {Senko}, \citenamefont {Vuletic}, \citenamefont {Greiner},\ and\ \citenamefont {Lukin}}]{Endres2016}%
  \BibitemOpen
  \bibfield  {author} {\bibinfo {author} {\bibfnamefont {Manuel}\ \bibnamefont {Endres}}, \bibinfo {author} {\bibfnamefont {Hannes}\ \bibnamefont {Bernien}}, \bibinfo {author} {\bibfnamefont {Alexander}\ \bibnamefont {Keesling}}, \bibinfo {author} {\bibfnamefont {Harry}\ \bibnamefont {Levine}}, \bibinfo {author} {\bibfnamefont {Eric~R.}\ \bibnamefont {Anschuetz}}, \bibinfo {author} {\bibfnamefont {Alexandre}\ \bibnamefont {Krajenbrink}}, \bibinfo {author} {\bibfnamefont {Crystal}\ \bibnamefont {Senko}}, \bibinfo {author} {\bibfnamefont {Vladan}\ \bibnamefont {Vuletic}}, \bibinfo {author} {\bibfnamefont {Markus}\ \bibnamefont {Greiner}}, \ and\ \bibinfo {author} {\bibfnamefont {Mikhail~D.}\ \bibnamefont {Lukin}},\ }\bibfield  {title} {\enquote {\bibinfo {title} {Atom-by-atom assembly of defect-free one-dimensional cold atom arrays},}\ }\href {https://doi.org/10.1126/science.aah3752} {\bibfield  {journal} {\bibinfo  {journal} {Science}\ }\textbf {\bibinfo {volume} {354}},\ \bibinfo {pages} {1024--1027}
  (\bibinfo {year} {2016})}\BibitemShut {NoStop}%
\bibitem [{\citenamefont {Manetsch}\ \emph {et~al.}(2024)\citenamefont {Manetsch}, \citenamefont {Nomura}, \citenamefont {Bataille}, \citenamefont {Leung}, \citenamefont {Lv},\ and\ \citenamefont {Endres}}]{Manetsch2024}%
  \BibitemOpen
  \bibfield  {author} {\bibinfo {author} {\bibfnamefont {Hannah~J.}\ \bibnamefont {Manetsch}}, \bibinfo {author} {\bibfnamefont {Gyohei}\ \bibnamefont {Nomura}}, \bibinfo {author} {\bibfnamefont {Elie}\ \bibnamefont {Bataille}}, \bibinfo {author} {\bibfnamefont {Kon~H.}\ \bibnamefont {Leung}}, \bibinfo {author} {\bibfnamefont {Xudong}\ \bibnamefont {Lv}}, \ and\ \bibinfo {author} {\bibfnamefont {Manuel}\ \bibnamefont {Endres}},\ }\bibfield  {title} {\enquote {\bibinfo {title} {A tweezer array with 6100 highly coherent atomic qubits},}\ }\href {https://doi.org/10.48550/arXiv.2403.12021} {\bibfield  {journal} {\bibinfo  {journal} {arXiv preprint}\ } (\bibinfo {year} {2024})}\BibitemShut {NoStop}%
\bibitem [{\citenamefont {Pause}\ \emph {et~al.}(2024)\citenamefont {Pause}, \citenamefont {Sturm}, \citenamefont {Mittenbühler}, \citenamefont {Amann}, \citenamefont {Preuschoff}, \citenamefont {Schäffner}, \citenamefont {Schlosser},\ and\ \citenamefont {Birkl}}]{Pause2024}%
  \BibitemOpen
  \bibfield  {author} {\bibinfo {author} {\bibfnamefont {Lars}\ \bibnamefont {Pause}}, \bibinfo {author} {\bibfnamefont {Lukas}\ \bibnamefont {Sturm}}, \bibinfo {author} {\bibfnamefont {Marcel}\ \bibnamefont {Mittenbühler}}, \bibinfo {author} {\bibfnamefont {Stephan}\ \bibnamefont {Amann}}, \bibinfo {author} {\bibfnamefont {Tilman}\ \bibnamefont {Preuschoff}}, \bibinfo {author} {\bibfnamefont {Dominik}\ \bibnamefont {Schäffner}}, \bibinfo {author} {\bibfnamefont {Malte}\ \bibnamefont {Schlosser}}, \ and\ \bibinfo {author} {\bibfnamefont {Gerhard}\ \bibnamefont {Birkl}},\ }\bibfield  {title} {\enquote {\bibinfo {title} {Supercharged two-dimensional tweezer array with more than 1000 atomic qubits},}\ }\href {https://doi.org/10.1364/OPTICA.513551} {\bibfield  {journal} {\bibinfo  {journal} {Optica}\ }\textbf {\bibinfo {volume} {11}},\ \bibinfo {pages} {222--226} (\bibinfo {year} {2024})}\BibitemShut {NoStop}%
\bibitem [{\citenamefont {Schlosser}\ \emph {et~al.}(2023)\citenamefont {Schlosser}, \citenamefont {Tichelmann}, \citenamefont {Schäffner}, \citenamefont {Ohl~de Mello}, \citenamefont {Hambach}, \citenamefont {Schütz},\ and\ \citenamefont {Birkl}}]{Schlosser2023}%
  \BibitemOpen
  \bibfield  {author} {\bibinfo {author} {\bibfnamefont {Malte}\ \bibnamefont {Schlosser}}, \bibinfo {author} {\bibfnamefont {Sascha}\ \bibnamefont {Tichelmann}}, \bibinfo {author} {\bibfnamefont {Dominik}\ \bibnamefont {Schäffner}}, \bibinfo {author} {\bibfnamefont {Daniel}\ \bibnamefont {Ohl~de Mello}}, \bibinfo {author} {\bibfnamefont {Moritz}\ \bibnamefont {Hambach}}, \bibinfo {author} {\bibfnamefont {Jan}\ \bibnamefont {Schütz}}, \ and\ \bibinfo {author} {\bibfnamefont {Gerhard}\ \bibnamefont {Birkl}},\ }\bibfield  {title} {\enquote {\bibinfo {title} {Scalable multilayer architecture of assembled single-atom qubit arrays in a three-dimensional talbot tweezer lattice},}\ }\href {https://doi.org/10.1103/PhysRevLett.130.180601} {\bibfield  {journal} {\bibinfo  {journal} {Phys. Rev. Lett.}\ }\textbf {\bibinfo {volume} {130}} (\bibinfo {year} {2023})}\BibitemShut {NoStop}%
\bibitem [{\citenamefont {Levine}\ \emph {et~al.}(2019)\citenamefont {Levine}, \citenamefont {Keesling}, \citenamefont {Semeghini}, \citenamefont {Omran}, \citenamefont {Wang}, \citenamefont {Ebadi}, \citenamefont {Bernien}, \citenamefont {Greiner}, \citenamefont {Vuletić}, \citenamefont {Pichler},\ and\ \citenamefont {Lukin}}]{Levine2019}%
  \BibitemOpen
  \bibfield  {author} {\bibinfo {author} {\bibfnamefont {Harry}\ \bibnamefont {Levine}}, \bibinfo {author} {\bibfnamefont {Alexander}\ \bibnamefont {Keesling}}, \bibinfo {author} {\bibfnamefont {Giulia}\ \bibnamefont {Semeghini}}, \bibinfo {author} {\bibfnamefont {Ahmed}\ \bibnamefont {Omran}}, \bibinfo {author} {\bibfnamefont {Tout~T.}\ \bibnamefont {Wang}}, \bibinfo {author} {\bibfnamefont {Sepehr}\ \bibnamefont {Ebadi}}, \bibinfo {author} {\bibfnamefont {Hannes}\ \bibnamefont {Bernien}}, \bibinfo {author} {\bibfnamefont {Markus}\ \bibnamefont {Greiner}}, \bibinfo {author} {\bibfnamefont {Vladan}\ \bibnamefont {Vuletić}}, \bibinfo {author} {\bibfnamefont {Hannes}\ \bibnamefont {Pichler}}, \ and\ \bibinfo {author} {\bibfnamefont {Mikhail~D.}\ \bibnamefont {Lukin}},\ }\bibfield  {title} {\enquote {\bibinfo {title} {Parallel implementation of high-fidelity multiqubit gates with neutral atoms},}\ }\href {https://doi.org/10.1103/PhysRevLett.123.170503} {\bibfield  {journal} {\bibinfo  {journal} {Nature}\
  }\textbf {\bibinfo {volume} {123}} (\bibinfo {year} {2019})}\BibitemShut {NoStop}%
\bibitem [{\citenamefont {Zhuo}\ \emph {et~al.}(2022)\citenamefont {Zhuo}, \citenamefont {Peng}, \citenamefont {Yuan}, \citenamefont {Yang-Yang}, \citenamefont {Xiao-Dong}, \citenamefont {Xiao}, \citenamefont {Min}, \citenamefont {Run-Bing}, \citenamefont {Jin}, \citenamefont {Liang}, ,\ and\ \citenamefont {Ming-Sheng}}]{Fu2022}%
  \BibitemOpen
  \bibfield  {author} {\bibinfo {author} {\bibfnamefont {Fu}~\bibnamefont {Zhuo}}, \bibinfo {author} {\bibfnamefont {Xu}~\bibnamefont {Peng}}, \bibinfo {author} {\bibfnamefont {Sun}\ \bibnamefont {Yuan}}, \bibinfo {author} {\bibfnamefont {Liu}\ \bibnamefont {Yang-Yang}}, \bibinfo {author} {\bibfnamefont {He}~\bibnamefont {Xiao-Dong}}, \bibinfo {author} {\bibfnamefont {Li}~\bibnamefont {Xiao}}, \bibinfo {author} {\bibfnamefont {Liu}\ \bibnamefont {Min}}, \bibinfo {author} {\bibfnamefont {Li}~\bibnamefont {Run-Bing}}, \bibinfo {author} {\bibfnamefont {Wang}\ \bibnamefont {Jin}}, \bibinfo {author} {\bibfnamefont {Liu}\ \bibnamefont {Liang}}, , \ and\ \bibinfo {author} {\bibfnamefont {Zhan}\ \bibnamefont {Ming-Sheng}},\ }\bibfield  {title} {\enquote {\bibinfo {title} {High-fidelity entanglement of neutral atoms via a {Rydberg}-mediated single-modulated-pulse controlled-phase gate},}\ }\href {https://doi.org/10.1103/PhysRevA.105.042430} {\bibfield  {journal} {\bibinfo  {journal} {Phys. Rev. A}\ }\textbf {\bibinfo
  {volume} {105}} (\bibinfo {year} {2022})}\BibitemShut {NoStop}%
\bibitem [{\citenamefont {Evered}\ \emph {et~al.}(2023)\citenamefont {Evered}, \citenamefont {Bluvstein}, \citenamefont {Kalinowski}, \citenamefont {Ebadi}, \citenamefont {Manovitz}, \citenamefont {Zhou}, \citenamefont {Li}, \citenamefont {Geim}, \citenamefont {Wang}, \citenamefont {Maskara}, \citenamefont {Levine}, \citenamefont {Semeghini}, \citenamefont {Greiner}, \citenamefont {Vuletić},\ and\ \citenamefont {Lukin}}]{Evered2023}%
  \BibitemOpen
  \bibfield  {author} {\bibinfo {author} {\bibfnamefont {Simon~J.}\ \bibnamefont {Evered}}, \bibinfo {author} {\bibfnamefont {Dolev}\ \bibnamefont {Bluvstein}}, \bibinfo {author} {\bibfnamefont {Marcin}\ \bibnamefont {Kalinowski}}, \bibinfo {author} {\bibfnamefont {Sepehr}\ \bibnamefont {Ebadi}}, \bibinfo {author} {\bibfnamefont {Tom}\ \bibnamefont {Manovitz}}, \bibinfo {author} {\bibfnamefont {Hengyun}\ \bibnamefont {Zhou}}, \bibinfo {author} {\bibfnamefont {Sophie~H.}\ \bibnamefont {Li}}, \bibinfo {author} {\bibfnamefont {Alexandra~A.}\ \bibnamefont {Geim}}, \bibinfo {author} {\bibfnamefont {Tout~T.}\ \bibnamefont {Wang}}, \bibinfo {author} {\bibfnamefont {Nishad}\ \bibnamefont {Maskara}}, \bibinfo {author} {\bibfnamefont {Harry}\ \bibnamefont {Levine}}, \bibinfo {author} {\bibfnamefont {Giulia}\ \bibnamefont {Semeghini}}, \bibinfo {author} {\bibfnamefont {Markus}\ \bibnamefont {Greiner}}, \bibinfo {author} {\bibfnamefont {Vladan}\ \bibnamefont {Vuletić}}, \ and\ \bibinfo {author} {\bibfnamefont
  {Mikhail~D.}\ \bibnamefont {Lukin}},\ }\bibfield  {title} {\enquote {\bibinfo {title} {High-fidelity parallel entangling gates on a neutral-atom quantum computer},}\ }\href {https://doi.org/10.1038/s41586-023-06481-y} {\bibfield  {journal} {\bibinfo  {journal} {Nature}\ }\textbf {\bibinfo {volume} {622}},\ \bibinfo {pages} {268--272} (\bibinfo {year} {2023})}\BibitemShut {NoStop}%
\bibitem [{\citenamefont {Bluvstein}\ \emph {et~al.}(2022)\citenamefont {Bluvstein}, \citenamefont {Levine}, \citenamefont {Semeghini}, \citenamefont {Wang}, \citenamefont {Ebadi}, \citenamefont {Kalinowski}, \citenamefont {Keesling}, \citenamefont {Maskara}, \citenamefont {Pichler}, \citenamefont {Greiner}, \citenamefont {Vuletić},\ and\ \citenamefont {Lukin}}]{Bluvstein2022}%
  \BibitemOpen
  \bibfield  {author} {\bibinfo {author} {\bibfnamefont {Dolev}\ \bibnamefont {Bluvstein}}, \bibinfo {author} {\bibfnamefont {Harry}\ \bibnamefont {Levine}}, \bibinfo {author} {\bibfnamefont {Giulia}\ \bibnamefont {Semeghini}}, \bibinfo {author} {\bibfnamefont {Tout~T.}\ \bibnamefont {Wang}}, \bibinfo {author} {\bibfnamefont {Sepehr}\ \bibnamefont {Ebadi}}, \bibinfo {author} {\bibfnamefont {Marcin}\ \bibnamefont {Kalinowski}}, \bibinfo {author} {\bibfnamefont {Alexander}\ \bibnamefont {Keesling}}, \bibinfo {author} {\bibfnamefont {Nishad}\ \bibnamefont {Maskara}}, \bibinfo {author} {\bibfnamefont {Hannes}\ \bibnamefont {Pichler}}, \bibinfo {author} {\bibfnamefont {Markus}\ \bibnamefont {Greiner}}, \bibinfo {author} {\bibfnamefont {Vladan}\ \bibnamefont {Vuletić}}, \ and\ \bibinfo {author} {\bibfnamefont {Mikhail~D.}\ \bibnamefont {Lukin}},\ }\bibfield  {title} {\enquote {\bibinfo {title} {A quantum processor based on coherent transport of entangled atom arrays},}\ }\href
  {https://doi.org/10.1038/s41586-022-04592-6} {\bibfield  {journal} {\bibinfo  {journal} {Nature}\ }\textbf {\bibinfo {volume} {604}},\ \bibinfo {pages} {451--456} (\bibinfo {year} {2022})}\BibitemShut {NoStop}%
\bibitem [{\citenamefont {Bluvstein}\ \emph {et~al.}(2023)\citenamefont {Bluvstein}, \citenamefont {Evered}, \citenamefont {Geim}, \citenamefont {Li}, \citenamefont {Zhou}, \citenamefont {Manovitz}, \citenamefont {Ebadi}, \citenamefont {Cain}, \citenamefont {Kalinowski}, \citenamefont {Hangleiter}, \citenamefont {Ataides}, \citenamefont {Maskara}, \citenamefont {Cong}, \citenamefont {Gao}, \citenamefont {Rodriguez}, \citenamefont {Karolyshyn}, \citenamefont {Semeghini}, \citenamefont {Gullans}, \citenamefont {Greiner}, \citenamefont {Vuletić},\ and\ \citenamefont {Lukin}}]{Bluvstein2023}%
  \BibitemOpen
  \bibfield  {author} {\bibinfo {author} {\bibfnamefont {Dolev}\ \bibnamefont {Bluvstein}}, \bibinfo {author} {\bibfnamefont {Simon~J.}\ \bibnamefont {Evered}}, \bibinfo {author} {\bibfnamefont {Alexandra~A.}\ \bibnamefont {Geim}}, \bibinfo {author} {\bibfnamefont {Sophie~H.}\ \bibnamefont {Li}}, \bibinfo {author} {\bibfnamefont {Hengyun}\ \bibnamefont {Zhou}}, \bibinfo {author} {\bibfnamefont {Tom}\ \bibnamefont {Manovitz}}, \bibinfo {author} {\bibfnamefont {Sepehr}\ \bibnamefont {Ebadi}}, \bibinfo {author} {\bibfnamefont {Madelyn}\ \bibnamefont {Cain}}, \bibinfo {author} {\bibfnamefont {Marcin}\ \bibnamefont {Kalinowski}}, \bibinfo {author} {\bibfnamefont {Dominik}\ \bibnamefont {Hangleiter}}, \bibinfo {author} {\bibfnamefont {J.~Pablo~Bonilla}\ \bibnamefont {Ataides}}, \bibinfo {author} {\bibfnamefont {Nishad}\ \bibnamefont {Maskara}}, \bibinfo {author} {\bibfnamefont {Iris}\ \bibnamefont {Cong}}, \bibinfo {author} {\bibfnamefont {Xun}\ \bibnamefont {Gao}}, \bibinfo {author} {\bibfnamefont {Pedro~Sales}\
  \bibnamefont {Rodriguez}}, \bibinfo {author} {\bibfnamefont {Thomas}\ \bibnamefont {Karolyshyn}}, \bibinfo {author} {\bibfnamefont {Giulia}\ \bibnamefont {Semeghini}}, \bibinfo {author} {\bibfnamefont {Michael~J.}\ \bibnamefont {Gullans}}, \bibinfo {author} {\bibfnamefont {Markus}\ \bibnamefont {Greiner}}, \bibinfo {author} {\bibfnamefont {Vladan}\ \bibnamefont {Vuletić}}, \ and\ \bibinfo {author} {\bibfnamefont {Mikhail~D.}\ \bibnamefont {Lukin}},\ }\bibfield  {title} {\enquote {\bibinfo {title} {Logical quantum processor based on reconfigurable atom arrays},}\ }\href {https://doi.org/10.1038/s41586-023-06927-3} {\bibfield  {journal} {\bibinfo  {journal} {Nature}\ }\textbf {\bibinfo {volume} {626}},\ \bibinfo {pages} {58--65} (\bibinfo {year} {2023})}\BibitemShut {NoStop}%
\bibitem [{\citenamefont {Gross}\ and\ \citenamefont {Bloch}(2017)}]{Gross2017}%
  \BibitemOpen
  \bibfield  {author} {\bibinfo {author} {\bibfnamefont {Christian}\ \bibnamefont {Gross}}\ and\ \bibinfo {author} {\bibfnamefont {Immanuel}\ \bibnamefont {Bloch}},\ }\bibfield  {title} {\enquote {\bibinfo {title} {Quantum simulations with ultracold atoms in optical lattices},}\ }\href {https://doi.org/10.1126/science.aal3837} {\bibfield  {journal} {\bibinfo  {journal} {Science}\ }\textbf {\bibinfo {volume} {357}},\ \bibinfo {pages} {995--1001} (\bibinfo {year} {2017})}\BibitemShut {NoStop}%
\bibitem [{\citenamefont {Wurtz}\ \emph {et~al.}(2023)\citenamefont {Wurtz}, \citenamefont {Bylinskii}, \citenamefont {Braverman}, \citenamefont {Amato-Grill}, \citenamefont {Cantu}, \citenamefont {Huber}, \citenamefont {Lukin}, \citenamefont {Liu}, \citenamefont {Weinberg}, \citenamefont {Long}, \citenamefont {Wang}, \citenamefont {Gemelke},\ and\ \citenamefont {Keesling}}]{Wurtz2023}%
  \BibitemOpen
  \bibfield  {author} {\bibinfo {author} {\bibfnamefont {Jonathan}\ \bibnamefont {Wurtz}}, \bibinfo {author} {\bibfnamefont {Alexei}\ \bibnamefont {Bylinskii}}, \bibinfo {author} {\bibfnamefont {Boris}\ \bibnamefont {Braverman}}, \bibinfo {author} {\bibfnamefont {Jesse}\ \bibnamefont {Amato-Grill}}, \bibinfo {author} {\bibfnamefont {Sergio~H.}\ \bibnamefont {Cantu}}, \bibinfo {author} {\bibfnamefont {Florian}\ \bibnamefont {Huber}}, \bibinfo {author} {\bibfnamefont {Alexander}\ \bibnamefont {Lukin}}, \bibinfo {author} {\bibfnamefont {Fangli}\ \bibnamefont {Liu}}, \bibinfo {author} {\bibfnamefont {Phillip}\ \bibnamefont {Weinberg}}, \bibinfo {author} {\bibfnamefont {John}\ \bibnamefont {Long}}, \bibinfo {author} {\bibfnamefont {Sheng-Tao}\ \bibnamefont {Wang}}, \bibinfo {author} {\bibfnamefont {Nathan}\ \bibnamefont {Gemelke}}, \ and\ \bibinfo {author} {\bibfnamefont {Alexander}\ \bibnamefont {Keesling}},\ }\bibfield  {title} {\enquote {\bibinfo {title} {Aquila: Quera's 256-qubit neutral-atom quantum
  computer},}\ }\href {https://doi.org/10.48550/arXiv.2306.11727} {\bibfield  {journal} {\bibinfo  {journal} {arXiv preprint}\ } (\bibinfo {year} {2023})}\BibitemShut {NoStop}%
\bibitem [{\citenamefont {Bernien}\ \emph {et~al.}(2017)\citenamefont {Bernien}, \citenamefont {Schwartz}, \citenamefont {Keesling}, \citenamefont {Levine}, \citenamefont {Omran}, \citenamefont {Pichler}, \citenamefont {Choi}, \citenamefont {Zibrov}, \citenamefont {Greiner}, \citenamefont {Vuletić},\ and\ \citenamefont {Lukin}}]{Bernien2017}%
  \BibitemOpen
  \bibfield  {author} {\bibinfo {author} {\bibfnamefont {H.}~\bibnamefont {Bernien}}, \bibinfo {author} {\bibfnamefont {S.}~\bibnamefont {Schwartz}}, \bibinfo {author} {\bibfnamefont {A.}~\bibnamefont {Keesling}}, \bibinfo {author} {\bibfnamefont {H.}~\bibnamefont {Levine}}, \bibinfo {author} {\bibfnamefont {A.}~\bibnamefont {Omran}}, \bibinfo {author} {\bibfnamefont {H.}~\bibnamefont {Pichler}}, \bibinfo {author} {\bibfnamefont {S.}~\bibnamefont {Choi}}, \bibinfo {author} {\bibfnamefont {A.~S.}\ \bibnamefont {Zibrov}}, \bibinfo {author} {\bibfnamefont {M.~En.~M.}\ \bibnamefont {Greiner}}, \bibinfo {author} {\bibfnamefont {V.}~\bibnamefont {Vuletić}}, \ and\ \bibinfo {author} {\bibfnamefont {M.~D.}\ \bibnamefont {Lukin}},\ }\bibfield  {title} {\enquote {\bibinfo {title} {Probing many-body dynamics on a 51-atom quantum simulator},}\ }\href {https://doi.org/10.1038/nature24622} {\bibfield  {journal} {\bibinfo  {journal} {Nature}\ }\textbf {\bibinfo {volume} {551}},\ \bibinfo {pages} {579–584} (\bibinfo
  {year} {2017})}\BibitemShut {NoStop}%
\bibitem [{\citenamefont {Keesling}\ \emph {et~al.}(2019)\citenamefont {Keesling}, \citenamefont {Omran}, \citenamefont {Levine}, \citenamefont {Bernien}, \citenamefont {Pichler}, \citenamefont {Choi}, \citenamefont {Samajdar}, \citenamefont {Schwartz}, \citenamefont {Silvi}, \citenamefont {Sachdev}, \citenamefont {Zoller}, \citenamefont {Endres}, \citenamefont {Greiner}, \citenamefont {Vuletić},\ and\ \citenamefont {Lukin}}]{Keesling2019}%
  \BibitemOpen
  \bibfield  {author} {\bibinfo {author} {\bibfnamefont {A.}~\bibnamefont {Keesling}}, \bibinfo {author} {\bibfnamefont {A.}~\bibnamefont {Omran}}, \bibinfo {author} {\bibfnamefont {H.}~\bibnamefont {Levine}}, \bibinfo {author} {\bibfnamefont {H.}~\bibnamefont {Bernien}}, \bibinfo {author} {\bibfnamefont {H.}~\bibnamefont {Pichler}}, \bibinfo {author} {\bibfnamefont {S.}~\bibnamefont {Choi}}, \bibinfo {author} {\bibfnamefont {R.}~\bibnamefont {Samajdar}}, \bibinfo {author} {\bibfnamefont {S.}~\bibnamefont {Schwartz}}, \bibinfo {author} {\bibfnamefont {P.}~\bibnamefont {Silvi}}, \bibinfo {author} {\bibfnamefont {S.}~\bibnamefont {Sachdev}}, \bibinfo {author} {\bibfnamefont {P.}~\bibnamefont {Zoller}}, \bibinfo {author} {\bibfnamefont {M.}~\bibnamefont {Endres}}, \bibinfo {author} {\bibfnamefont {M.}~\bibnamefont {Greiner}}, \bibinfo {author} {\bibfnamefont {V.}~\bibnamefont {Vuletić}}, \ and\ \bibinfo {author} {\bibfnamefont {M.~D.}\ \bibnamefont {Lukin}},\ }\bibfield  {title} {\enquote {\bibinfo {title}
  {Quantum {Kibble–Zurek} mechanism and critical dynamics on a programmable {Rydberg} simulator},}\ }\href {https://doi.org/10.1038/s41586-019-1070-1} {\bibfield  {journal} {\bibinfo  {journal} {Nature}\ }\textbf {\bibinfo {volume} {568}},\ \bibinfo {pages} {207--211} (\bibinfo {year} {2019})}\BibitemShut {NoStop}%
\bibitem [{\citenamefont {De~Léséleuc}\ \emph {et~al.}(2019)\citenamefont {De~Léséleuc}, \citenamefont {Lienhard}, \citenamefont {Scholl}, \citenamefont {Barredo}, \citenamefont {Weber}, \citenamefont {Lang}, \citenamefont {Büchler}, \citenamefont {Lahaye},\ and\ \citenamefont {Browaeys}}]{Leseleuc2019}%
  \BibitemOpen
  \bibfield  {author} {\bibinfo {author} {\bibfnamefont {Sylvain}\ \bibnamefont {De~Léséleuc}}, \bibinfo {author} {\bibfnamefont {Vincent}\ \bibnamefont {Lienhard}}, \bibinfo {author} {\bibfnamefont {Pascal}\ \bibnamefont {Scholl}}, \bibinfo {author} {\bibfnamefont {Daniel}\ \bibnamefont {Barredo}}, \bibinfo {author} {\bibfnamefont {Sebastian}\ \bibnamefont {Weber}}, \bibinfo {author} {\bibfnamefont {Nicolai}\ \bibnamefont {Lang}}, \bibinfo {author} {\bibfnamefont {Hans~Peter}\ \bibnamefont {Büchler}}, \bibinfo {author} {\bibfnamefont {Thierry}\ \bibnamefont {Lahaye}}, \ and\ \bibinfo {author} {\bibfnamefont {Antoine}\ \bibnamefont {Browaeys}},\ }\bibfield  {title} {\enquote {\bibinfo {title} {Observation of a symmetry-protected topological phase of interacting bosons with {Rydberg} atoms},}\ }\href {https://doi.org/10.1126/science.aav9105} {\bibfield  {journal} {\bibinfo  {journal} {Science}\ }\textbf {\bibinfo {volume} {365}},\ \bibinfo {pages} {775--780} (\bibinfo {year} {2019})}\BibitemShut {NoStop}%
\bibitem [{\citenamefont {Scholl}\ \emph {et~al.}(2021)\citenamefont {Scholl}, \citenamefont {Schuler}, \citenamefont {Williams}, \citenamefont {Eberharter}, \citenamefont {Barredo}, \citenamefont {Schymik}, \citenamefont {Lienhard}, \citenamefont {Henry}, \citenamefont {Lang}, \citenamefont {Lahaye}, \citenamefont {Läuchli},\ and\ \citenamefont {Browaeys}}]{Scholl2021}%
  \BibitemOpen
  \bibfield  {author} {\bibinfo {author} {\bibfnamefont {Pascal}\ \bibnamefont {Scholl}}, \bibinfo {author} {\bibfnamefont {Michael}\ \bibnamefont {Schuler}}, \bibinfo {author} {\bibfnamefont {Hannah~J.}\ \bibnamefont {Williams}}, \bibinfo {author} {\bibfnamefont {Alexander~A.}\ \bibnamefont {Eberharter}}, \bibinfo {author} {\bibfnamefont {Daniel}\ \bibnamefont {Barredo}}, \bibinfo {author} {\bibfnamefont {Kai-Niklas}\ \bibnamefont {Schymik}}, \bibinfo {author} {\bibfnamefont {Vincent}\ \bibnamefont {Lienhard}}, \bibinfo {author} {\bibfnamefont {Louis-Paul}\ \bibnamefont {Henry}}, \bibinfo {author} {\bibfnamefont {Thomas~C.}\ \bibnamefont {Lang}}, \bibinfo {author} {\bibfnamefont {Thierry}\ \bibnamefont {Lahaye}}, \bibinfo {author} {\bibfnamefont {Andreas~M.}\ \bibnamefont {Läuchli}}, \ and\ \bibinfo {author} {\bibfnamefont {Antoine}\ \bibnamefont {Browaeys}},\ }\bibfield  {title} {\enquote {\bibinfo {title} {Quantum simulation of 2d antiferromagnets with hundreds of {Rydberg} atoms},}\ }\href
  {https://doi.org/10.1038/s41586-021-03585-1} {\bibfield  {journal} {\bibinfo  {journal} {Nature}\ }\textbf {\bibinfo {volume} {595}},\ \bibinfo {pages} {233–238} (\bibinfo {year} {2021})}\BibitemShut {NoStop}%
\bibitem [{\citenamefont {Ebadi}\ \emph {et~al.}(2021)\citenamefont {Ebadi}, \citenamefont {Wang}, \citenamefont {Levine}, \citenamefont {Keesling}, \citenamefont {Semeghini}, \citenamefont {Omran}, \citenamefont {Bluvstein}, \citenamefont {Samajdar}, \citenamefont {Pichler}, \citenamefont {Ho}, \citenamefont {Choi}, \citenamefont {Sachdev}, \citenamefont {Greiner}, \citenamefont {Vuletić},\ and\ \citenamefont {Lukin}}]{Ebadi2021}%
  \BibitemOpen
  \bibfield  {author} {\bibinfo {author} {\bibfnamefont {Sepehr}\ \bibnamefont {Ebadi}}, \bibinfo {author} {\bibfnamefont {Tout~T.}\ \bibnamefont {Wang}}, \bibinfo {author} {\bibfnamefont {Harry}\ \bibnamefont {Levine}}, \bibinfo {author} {\bibfnamefont {Alexander}\ \bibnamefont {Keesling}}, \bibinfo {author} {\bibfnamefont {Giulia}\ \bibnamefont {Semeghini}}, \bibinfo {author} {\bibfnamefont {Ahmed}\ \bibnamefont {Omran}}, \bibinfo {author} {\bibfnamefont {Dolev}\ \bibnamefont {Bluvstein}}, \bibinfo {author} {\bibfnamefont {Rhine}\ \bibnamefont {Samajdar}}, \bibinfo {author} {\bibfnamefont {Hannes}\ \bibnamefont {Pichler}}, \bibinfo {author} {\bibfnamefont {Wen~Wei}\ \bibnamefont {Ho}}, \bibinfo {author} {\bibfnamefont {Soonwon}\ \bibnamefont {Choi}}, \bibinfo {author} {\bibfnamefont {Subir}\ \bibnamefont {Sachdev}}, \bibinfo {author} {\bibfnamefont {Markus}\ \bibnamefont {Greiner}}, \bibinfo {author} {\bibfnamefont {Vladan}\ \bibnamefont {Vuletić}}, \ and\ \bibinfo {author} {\bibfnamefont {Mikhail~D.}\
  \bibnamefont {Lukin}},\ }\bibfield  {title} {\enquote {\bibinfo {title} {Quantum phases of matter on a 256-atom programmable quantum simulator},}\ }\href {https://doi.org/10.1038/s41586-021-03582-4} {\bibfield  {journal} {\bibinfo  {journal} {Nature}\ }\textbf {\bibinfo {volume} {595}},\ \bibinfo {pages} {227–232} (\bibinfo {year} {2021})}\BibitemShut {NoStop}%
\bibitem [{\citenamefont {Zhang}\ \emph {et~al.}(2024)\citenamefont {Zhang}, \citenamefont {Cantú}, \citenamefont {Liu}, \citenamefont {Bylinskii}, \citenamefont {Braverman}, \citenamefont {Huber}, \citenamefont {Amato-Grill}, \citenamefont {Lukin}, \citenamefont {Gemelke}, \citenamefont {Keesling}, \citenamefont {Wang}, \citenamefont {Meurice},\ and\ \citenamefont {Tsai}}]{Zhang2024}%
  \BibitemOpen
  \bibfield  {author} {\bibinfo {author} {\bibfnamefont {Jin}\ \bibnamefont {Zhang}}, \bibinfo {author} {\bibfnamefont {Sergio~H.}\ \bibnamefont {Cantú}}, \bibinfo {author} {\bibfnamefont {Fangli}\ \bibnamefont {Liu}}, \bibinfo {author} {\bibfnamefont {Alexei}\ \bibnamefont {Bylinskii}}, \bibinfo {author} {\bibfnamefont {Boris}\ \bibnamefont {Braverman}}, \bibinfo {author} {\bibfnamefont {Florian}\ \bibnamefont {Huber}}, \bibinfo {author} {\bibfnamefont {Jesse}\ \bibnamefont {Amato-Grill}}, \bibinfo {author} {\bibfnamefont {Alexander}\ \bibnamefont {Lukin}}, \bibinfo {author} {\bibfnamefont {Nathan}\ \bibnamefont {Gemelke}}, \bibinfo {author} {\bibfnamefont {Alexander}\ \bibnamefont {Keesling}}, \bibinfo {author} {\bibfnamefont {Sheng-Tao}\ \bibnamefont {Wang}}, \bibinfo {author} {\bibfnamefont {Y.}~\bibnamefont {Meurice}}, \ and\ \bibinfo {author} {\bibfnamefont {S.-W.}\ \bibnamefont {Tsai}},\ }\bibfield  {title} {\enquote {\bibinfo {title} {Probing quantum floating phases in {Rydberg} atom arrays},}\
  }\href {https://doi.org/10.48550/arXiv.2401.08087} {\bibfield  {journal} {\bibinfo  {journal} {arXiv preprint 2401.08087}\ } (\bibinfo {year} {2024})}\BibitemShut {NoStop}%
\bibitem [{\citenamefont {Amico}\ \emph {et~al.}(2008)\citenamefont {Amico}, \citenamefont {Fazio}, \citenamefont {Osterloh},\ and\ \citenamefont {Vedral}}]{Amico2008}%
  \BibitemOpen
  \bibfield  {author} {\bibinfo {author} {\bibfnamefont {Luigi}\ \bibnamefont {Amico}}, \bibinfo {author} {\bibfnamefont {Rosario}\ \bibnamefont {Fazio}}, \bibinfo {author} {\bibfnamefont {Andreas}\ \bibnamefont {Osterloh}}, \ and\ \bibinfo {author} {\bibfnamefont {Vlatko}\ \bibnamefont {Vedral}},\ }\bibfield  {title} {\enquote {\bibinfo {title} {Entanglement in many-body systems},}\ }\href {https://doi.org/10.1103/RevModPhys.80.517} {\bibfield  {journal} {\bibinfo  {journal} {Rev. Mod. Phys.}\ }\textbf {\bibinfo {volume} {80}} (\bibinfo {year} {2008})}\BibitemShut {NoStop}%
\bibitem [{\citenamefont {Baccari}\ \emph {et~al.}(2017)\citenamefont {Baccari}, \citenamefont {Cavalcanti}, \citenamefont {Wittek},\ and\ \citenamefont {Acín}}]{Baccari2017}%
  \BibitemOpen
  \bibfield  {author} {\bibinfo {author} {\bibfnamefont {Flavio}\ \bibnamefont {Baccari}}, \bibinfo {author} {\bibfnamefont {Daniel}\ \bibnamefont {Cavalcanti}}, \bibinfo {author} {\bibfnamefont {Peter}\ \bibnamefont {Wittek}}, \ and\ \bibinfo {author} {\bibfnamefont {Antonio}\ \bibnamefont {Acín}},\ }\bibfield  {title} {\enquote {\bibinfo {title} {Efficient device-independent entanglement detection for multipartite systems},}\ }\href {https://doi.org/10.1103/PhysRevX.7.021042} {\bibfield  {journal} {\bibinfo  {journal} {Phys. Rev. X}\ }\textbf {\bibinfo {volume} {7}} (\bibinfo {year} {2017})}\BibitemShut {NoStop}%
\bibitem [{\citenamefont {Friis}\ \emph {et~al.}(2018)\citenamefont {Friis}, \citenamefont {Vitagliano}, \citenamefont {Malik},\ and\ \citenamefont {Marcus}}]{Friis2018}%
  \BibitemOpen
  \bibfield  {author} {\bibinfo {author} {\bibfnamefont {Nicolai}\ \bibnamefont {Friis}}, \bibinfo {author} {\bibfnamefont {Giuseppe}\ \bibnamefont {Vitagliano}}, \bibinfo {author} {\bibfnamefont {Mehul}\ \bibnamefont {Malik}}, \ and\ \bibinfo {author} {\bibfnamefont {Huber}\ \bibnamefont {Marcus}},\ }\bibfield  {title} {\enquote {\bibinfo {title} {Entanglement certification from theory to experiment},}\ }\href {https://doi.org/10.1038/s42254-018-0003-5} {\bibfield  {journal} {\bibinfo  {journal} {Nature Reviews Physics}\ }\textbf {\bibinfo {volume} {1}},\ \bibinfo {pages} {72–87} (\bibinfo {year} {2018})}\BibitemShut {NoStop}%
\bibitem [{\citenamefont {Daley}\ \emph {et~al.}(2012)\citenamefont {Daley}, \citenamefont {Pichler}, \citenamefont {Schachenmayer},\ and\ \citenamefont {Zoller}}]{Daley2012}%
  \BibitemOpen
  \bibfield  {author} {\bibinfo {author} {\bibfnamefont {Andrew~J.}\ \bibnamefont {Daley}}, \bibinfo {author} {\bibfnamefont {Hannes}\ \bibnamefont {Pichler}}, \bibinfo {author} {\bibfnamefont {Johannes}\ \bibnamefont {Schachenmayer}}, \ and\ \bibinfo {author} {\bibfnamefont {Peter}\ \bibnamefont {Zoller}},\ }\bibfield  {title} {\enquote {\bibinfo {title} {Measuring entanglement growth in quench dynamics of bosons in an optical lattice},}\ }\href {https://doi.org/10.1103/PhysRevLett.109.020505} {\bibfield  {journal} {\bibinfo  {journal} {Phys. Rev. Lett.}\ }\textbf {\bibinfo {volume} {109}} (\bibinfo {year} {2012})}\BibitemShut {NoStop}%
\bibitem [{\citenamefont {Islam}\ \emph {et~al.}(2015)\citenamefont {Islam}, \citenamefont {Ma}, \citenamefont {Preiss}, \citenamefont {Tai}, \citenamefont {Lukin}, \citenamefont {Rispoli},\ and\ \citenamefont {Greiner}}]{Islam2015}%
  \BibitemOpen
  \bibfield  {author} {\bibinfo {author} {\bibfnamefont {Rajibul}\ \bibnamefont {Islam}}, \bibinfo {author} {\bibfnamefont {Ruichao}\ \bibnamefont {Ma}}, \bibinfo {author} {\bibfnamefont {Philipp~M.}\ \bibnamefont {Preiss}}, \bibinfo {author} {\bibfnamefont {M.~Eric}\ \bibnamefont {Tai}}, \bibinfo {author} {\bibfnamefont {Alexander}\ \bibnamefont {Lukin}}, \bibinfo {author} {\bibfnamefont {Matthew}\ \bibnamefont {Rispoli}}, \ and\ \bibinfo {author} {\bibfnamefont {Markus}\ \bibnamefont {Greiner}},\ }\bibfield  {title} {\enquote {\bibinfo {title} {Measuring entanglement entropy in a quantum many-body system},}\ }\href {https://doi.org/10.1038/nature15750} {\bibfield  {journal} {\bibinfo  {journal} {Nature}\ }\textbf {\bibinfo {volume} {528}},\ \bibinfo {pages} {77–83} (\bibinfo {year} {2015})}\BibitemShut {NoStop}%
\bibitem [{\citenamefont {Elben}\ \emph {et~al.}(2023)\citenamefont {Elben}, \citenamefont {Flammia}, \citenamefont {Huang}, \citenamefont {Kueng}, \citenamefont {Preskill}, \citenamefont {Vermersch},\ and\ \citenamefont {Zoller}}]{Elben2023}%
  \BibitemOpen
  \bibfield  {author} {\bibinfo {author} {\bibfnamefont {Andreas}\ \bibnamefont {Elben}}, \bibinfo {author} {\bibfnamefont {Steven~T.}\ \bibnamefont {Flammia}}, \bibinfo {author} {\bibfnamefont {Hsin-Yuan}\ \bibnamefont {Huang}}, \bibinfo {author} {\bibfnamefont {Richard}\ \bibnamefont {Kueng}}, \bibinfo {author} {\bibfnamefont {John}\ \bibnamefont {Preskill}}, \bibinfo {author} {\bibfnamefont {Benoît}\ \bibnamefont {Vermersch}}, \ and\ \bibinfo {author} {\bibfnamefont {Peter}\ \bibnamefont {Zoller}},\ }\bibfield  {title} {\enquote {\bibinfo {title} {The randomized measurement toolbox},}\ }\href {https://doi.org/10.1038/s42254-022-00535-2} {\bibfield  {journal} {\bibinfo  {journal} {Nature Reviews Physics}\ }\textbf {\bibinfo {volume} {5}},\ \bibinfo {pages} {9–24} (\bibinfo {year} {2023})}\BibitemShut {NoStop}%
\bibitem [{\citenamefont {Notarnicola}\ \emph {et~al.}(2023)\citenamefont {Notarnicola}, \citenamefont {Elben}, \citenamefont {Lahaye}, \citenamefont {Browaeys}, \citenamefont {Montangero},\ and\ \citenamefont {Vermersch}}]{Notarnicola2023}%
  \BibitemOpen
  \bibfield  {author} {\bibinfo {author} {\bibfnamefont {Simone}\ \bibnamefont {Notarnicola}}, \bibinfo {author} {\bibfnamefont {Andreas}\ \bibnamefont {Elben}}, \bibinfo {author} {\bibfnamefont {Thierry}\ \bibnamefont {Lahaye}}, \bibinfo {author} {\bibfnamefont {Antoine}\ \bibnamefont {Browaeys}}, \bibinfo {author} {\bibfnamefont {Simone}\ \bibnamefont {Montangero}}, \ and\ \bibinfo {author} {\bibfnamefont {Benoît}\ \bibnamefont {Vermersch}},\ }\bibfield  {title} {\enquote {\bibinfo {title} {A randomized measurement toolbox for an interacting {Rydberg}-atom quantum simulator},}\ }\href {https://doi.org/10.1088/1367-2630/acfcd3} {\bibfield  {journal} {\bibinfo  {journal} {New Journal of Physics}\ }\textbf {\bibinfo {volume} {25}} (\bibinfo {year} {2023})}\BibitemShut {NoStop}%
\bibitem [{\citenamefont {Elben}\ \emph {et~al.}(2020)\citenamefont {Elben}, \citenamefont {Kueng}, \citenamefont {Huang}, \citenamefont {van Bijnen}, \citenamefont {Kokail}, \citenamefont {Dalmonte}, \citenamefont {Calabrese}, \citenamefont {Kraus}, \citenamefont {Preskill}, \citenamefont {Zoller},\ and\ \citenamefont {Vermersch}}]{Elben2020}%
  \BibitemOpen
  \bibfield  {author} {\bibinfo {author} {\bibfnamefont {Andreas}\ \bibnamefont {Elben}}, \bibinfo {author} {\bibfnamefont {Richard}\ \bibnamefont {Kueng}}, \bibinfo {author} {\bibfnamefont {Hsin-Yuan~(Robert)}\ \bibnamefont {Huang}}, \bibinfo {author} {\bibfnamefont {Rick}\ \bibnamefont {van Bijnen}}, \bibinfo {author} {\bibfnamefont {Christian}\ \bibnamefont {Kokail}}, \bibinfo {author} {\bibfnamefont {Marcello}\ \bibnamefont {Dalmonte}}, \bibinfo {author} {\bibfnamefont {Pasquale}\ \bibnamefont {Calabrese}}, \bibinfo {author} {\bibfnamefont {Barbara}\ \bibnamefont {Kraus}}, \bibinfo {author} {\bibfnamefont {John}\ \bibnamefont {Preskill}}, \bibinfo {author} {\bibfnamefont {Peter}\ \bibnamefont {Zoller}}, \ and\ \bibinfo {author} {\bibfnamefont {Benoît}\ \bibnamefont {Vermersch}},\ }\bibfield  {title} {\enquote {\bibinfo {title} {Mixed-state entanglement from local randomized measurements},}\ }\href {https://doi.org/10.1103/PhysRevLett.125.200501} {\bibfield  {journal} {\bibinfo  {journal} {Phys. Rev.
  Lett.}\ }\textbf {\bibinfo {volume} {125}} (\bibinfo {year} {2020})}\BibitemShut {NoStop}%
\bibitem [{\citenamefont {Eisert}\ \emph {et~al.}(2010)\citenamefont {Eisert}, \citenamefont {Cramer},\ and\ \citenamefont {Plenio}}]{Eisert2010}%
  \BibitemOpen
  \bibfield  {author} {\bibinfo {author} {\bibfnamefont {J.}~\bibnamefont {Eisert}}, \bibinfo {author} {\bibfnamefont {M.}~\bibnamefont {Cramer}}, \ and\ \bibinfo {author} {\bibfnamefont {M.~B.}\ \bibnamefont {Plenio}},\ }\bibfield  {title} {\enquote {\bibinfo {title} {Colloquium: Area laws for the entanglement entropy},}\ }\href {https://doi.org/10.1103/RevModPhys.82.277} {\bibfield  {journal} {\bibinfo  {journal} {Rev. Mod. Phys.}\ }\textbf {\bibinfo {volume} {82}} (\bibinfo {year} {2010})}\BibitemShut {NoStop}%
\bibitem [{\citenamefont {Laflorencie}(2016)}]{Laflorencie2016}%
  \BibitemOpen
  \bibfield  {author} {\bibinfo {author} {\bibfnamefont {Nicolas}\ \bibnamefont {Laflorencie}},\ }\bibfield  {title} {\enquote {\bibinfo {title} {Quantum entanglement in condensed matter systems},}\ }\href {https://doi.org/10.1016/j.physrep.2016.06.008} {\bibfield  {journal} {\bibinfo  {journal} {Rev. Mod. Phys.}\ }\textbf {\bibinfo {volume} {646}},\ \bibinfo {pages} {1--59} (\bibinfo {year} {2016})}\BibitemShut {NoStop}%
\bibitem [{\citenamefont {Bennett}\ \emph {et~al.}(1996{\natexlab{a}})\citenamefont {Bennett}, \citenamefont {Bernstein}, \citenamefont {Popescu},\ and\ \citenamefont {Schumacher}}]{Bennet1996}%
  \BibitemOpen
  \bibfield  {author} {\bibinfo {author} {\bibfnamefont {Charles~H.}\ \bibnamefont {Bennett}}, \bibinfo {author} {\bibfnamefont {Herbert~J.}\ \bibnamefont {Bernstein}}, \bibinfo {author} {\bibfnamefont {Sandu}\ \bibnamefont {Popescu}}, \ and\ \bibinfo {author} {\bibfnamefont {Benjamin}\ \bibnamefont {Schumacher}},\ }\bibfield  {title} {\enquote {\bibinfo {title} {Concentrating partial entanglement by local operations},}\ }\href {https://doi.org/10.1103/physreva.53.2046} {\bibfield  {journal} {\bibinfo  {journal} {Phys. Rev. A}\ }\textbf {\bibinfo {volume} {53}} (\bibinfo {year} {1996}{\natexlab{a}})}\BibitemShut {NoStop}%
\bibitem [{\citenamefont {Cui}\ \emph {et~al.}(2013)\citenamefont {Cui}, \citenamefont {Amico}, \citenamefont {Fan}, \citenamefont {Gu}, \citenamefont {Hamma},\ and\ \citenamefont {Vedral}}]{Cui2013}%
  \BibitemOpen
  \bibfield  {author} {\bibinfo {author} {\bibfnamefont {Jian}\ \bibnamefont {Cui}}, \bibinfo {author} {\bibfnamefont {Luigi}\ \bibnamefont {Amico}}, \bibinfo {author} {\bibfnamefont {Heng}\ \bibnamefont {Fan}}, \bibinfo {author} {\bibfnamefont {Mile}\ \bibnamefont {Gu}}, \bibinfo {author} {\bibfnamefont {Alioscia}\ \bibnamefont {Hamma}}, \ and\ \bibinfo {author} {\bibfnamefont {Vlatko}\ \bibnamefont {Vedral}},\ }\bibfield  {title} {\enquote {\bibinfo {title} {Local characterization of one-dimensional topologically ordered states},}\ }\href {https://doi.org/10.1103/PhysRevB.88.125117} {\bibfield  {journal} {\bibinfo  {journal} {Phys. Rev. B}\ }\textbf {\bibinfo {volume} {88}} (\bibinfo {year} {2013})}\BibitemShut {NoStop}%
\bibitem [{\citenamefont {Bruß}(2002)}]{Brus2002}%
  \BibitemOpen
  \bibfield  {author} {\bibinfo {author} {\bibfnamefont {Dagmar}\ \bibnamefont {Bruß}},\ }\bibfield  {title} {\enquote {\bibinfo {title} {Characterizing entanglement},}\ }\href {https://doi.org/10.1063/1.1494474} {\bibfield  {journal} {\bibinfo  {journal} {J. Math. Phys.}\ }\textbf {\bibinfo {volume} {43}},\ \bibinfo {pages} {4237–4251} (\bibinfo {year} {2002})}\BibitemShut {NoStop}%
\bibitem [{\citenamefont {Plenio}\ and\ \citenamefont {Virmani}(2007)}]{Plenio2007}%
  \BibitemOpen
  \bibfield  {author} {\bibinfo {author} {\bibfnamefont {Martin~B.}\ \bibnamefont {Plenio}}\ and\ \bibinfo {author} {\bibfnamefont {Shashank}\ \bibnamefont {Virmani}},\ }\bibfield  {title} {\enquote {\bibinfo {title} {An introduction to entanglement measures},}\ }\href {https://doi.org/10.26421/QIC7.1-2-1} {\bibfield  {journal} {\bibinfo  {journal} {Quant. Inf. Comput.}\ }\textbf {\bibinfo {volume} {7}},\ \bibinfo {pages} {1--51} (\bibinfo {year} {2007})}\BibitemShut {NoStop}%
\bibitem [{\citenamefont {Bennett}\ \emph {et~al.}(1996{\natexlab{b}})\citenamefont {Bennett}, \citenamefont {DiVincenzo}, \citenamefont {Smolin},\ and\ \citenamefont {Wootters}}]{Bennet1996eof}%
  \BibitemOpen
  \bibfield  {author} {\bibinfo {author} {\bibfnamefont {Charles~H.}\ \bibnamefont {Bennett}}, \bibinfo {author} {\bibfnamefont {David~P.}\ \bibnamefont {DiVincenzo}}, \bibinfo {author} {\bibfnamefont {John~A.}\ \bibnamefont {Smolin}}, \ and\ \bibinfo {author} {\bibfnamefont {William~K.}\ \bibnamefont {Wootters}},\ }\bibfield  {title} {\enquote {\bibinfo {title} {Mixed-state entanglement and quantum error correction},}\ }\href {https://doi.org/10.1103/PhysRevA.54.3824} {\bibfield  {journal} {\bibinfo  {journal} {Phys. Rev. A}\ }\textbf {\bibinfo {volume} {54}} (\bibinfo {year} {1996}{\natexlab{b}})}\BibitemShut {NoStop}%
\bibitem [{\citenamefont {Arceci}\ \emph {et~al.}(2021)\citenamefont {Arceci}, \citenamefont {Silvi},\ and\ \citenamefont {Montangero}}]{TTOpaper}%
  \BibitemOpen
  \bibfield  {author} {\bibinfo {author} {\bibfnamefont {L.}~\bibnamefont {Arceci}}, \bibinfo {author} {\bibfnamefont {P.}~\bibnamefont {Silvi}}, \ and\ \bibinfo {author} {\bibfnamefont {S.}~\bibnamefont {Montangero}},\ }\bibfield  {title} {\enquote {\bibinfo {title} {Entanglement of formation of mixed many-body quantum states via tree tensor networks},}\ }\href {https://doi.org/10.1103/PhysRevLett.128.040501} {\bibfield  {journal} {\bibinfo  {journal} {Phys. Rev. Lett.}\ }\textbf {\bibinfo {volume} {128}} (\bibinfo {year} {2021})}\BibitemShut {NoStop}%
\bibitem [{\citenamefont {Werner}\ \emph {et~al.}(2016)\citenamefont {Werner}, \citenamefont {Jaschke}, \citenamefont {Silvi}, \citenamefont {Kliesch}, \citenamefont {Calarco}, \citenamefont {Eisert},\ and\ \citenamefont {S.}}]{Werner2016}%
  \BibitemOpen
  \bibfield  {author} {\bibinfo {author} {\bibfnamefont {A.~H.}\ \bibnamefont {Werner}}, \bibinfo {author} {\bibfnamefont {D.}~\bibnamefont {Jaschke}}, \bibinfo {author} {\bibfnamefont {P.}~\bibnamefont {Silvi}}, \bibinfo {author} {\bibfnamefont {M.}~\bibnamefont {Kliesch}}, \bibinfo {author} {\bibfnamefont {T.}~\bibnamefont {Calarco}}, \bibinfo {author} {\bibfnamefont {J.}~\bibnamefont {Eisert}}, \ and\ \bibinfo {author} {\bibfnamefont {Montangero}\ \bibnamefont {S.}},\ }\bibfield  {title} {\enquote {\bibinfo {title} {Positive tensor network approach for simulating open quantum many-body systems},}\ }\href {https://doi.org/10.1103/PhysRevLett.116.237201} {\bibfield  {journal} {\bibinfo  {journal} {Phys. Rev. Lett.}\ }\textbf {\bibinfo {volume} {116}} (\bibinfo {year} {2016})}\BibitemShut {NoStop}%
\bibitem [{\citenamefont {Schollwöck}(2011)}]{Schollwock2011}%
  \BibitemOpen
  \bibfield  {author} {\bibinfo {author} {\bibfnamefont {Ulrich}\ \bibnamefont {Schollwöck}},\ }\bibfield  {title} {\enquote {\bibinfo {title} {The density-matrix renormalization group in the age of matrix product states},}\ }\href {https://doi.org/10.1016/j.aop.2010.09.012} {\bibfield  {journal} {\bibinfo  {journal} {Annals of Physics}\ }\textbf {\bibinfo {volume} {326}},\ \bibinfo {pages} {96--192} (\bibinfo {year} {2011})}\BibitemShut {NoStop}%
\bibitem [{\citenamefont {Vidal}\ and\ \citenamefont {Werner}(2002)}]{Vidal2002}%
  \BibitemOpen
  \bibfield  {author} {\bibinfo {author} {\bibfnamefont {G.}~\bibnamefont {Vidal}}\ and\ \bibinfo {author} {\bibfnamefont {R.~F.}\ \bibnamefont {Werner}},\ }\bibfield  {title} {\enquote {\bibinfo {title} {Computable measure of entanglement},}\ }\href {https://doi.org/10.1103/PhysRevA.65.032314} {\bibfield  {journal} {\bibinfo  {journal} {Phys. Rev. A}\ }\textbf {\bibinfo {volume} {65}} (\bibinfo {year} {2002})}\BibitemShut {NoStop}%
\bibitem [{\citenamefont {Wichterich}\ \emph {et~al.}(2010)\citenamefont {Wichterich}, \citenamefont {Vidal},\ and\ \citenamefont {Bose}}]{Wichterich2010}%
  \BibitemOpen
  \bibfield  {author} {\bibinfo {author} {\bibfnamefont {Hannu}\ \bibnamefont {Wichterich}}, \bibinfo {author} {\bibfnamefont {Julien}\ \bibnamefont {Vidal}}, \ and\ \bibinfo {author} {\bibfnamefont {Sougato}\ \bibnamefont {Bose}},\ }\bibfield  {title} {\enquote {\bibinfo {title} {Universality of the negativity in the {Lipkin}-{Meshkov}-{Glick} model},}\ }\href {https://doi.org/10.1103/PhysRevA.81.032311} {\bibfield  {journal} {\bibinfo  {journal} {Phys. Rev. A}\ }\textbf {\bibinfo {volume} {81}} (\bibinfo {year} {2010})}\BibitemShut {NoStop}%
\bibitem [{\citenamefont {Calabrese}\ \emph {et~al.}(2014)\citenamefont {Calabrese}, \citenamefont {Cardy},\ and\ \citenamefont {Tonni}}]{Calabrese2014}%
  \BibitemOpen
  \bibfield  {author} {\bibinfo {author} {\bibfnamefont {Pasquale}\ \bibnamefont {Calabrese}}, \bibinfo {author} {\bibfnamefont {John}\ \bibnamefont {Cardy}}, \ and\ \bibinfo {author} {\bibfnamefont {Erik}\ \bibnamefont {Tonni}},\ }\bibfield  {title} {\enquote {\bibinfo {title} {Finite temperature entanglement negativity in conformal field theory},}\ }\href {https://doi.org/10.1088/1751-8113/48/1/015006} {\bibfield  {journal} {\bibinfo  {journal} {J. Phys. A: Math. Theor.}\ }\textbf {\bibinfo {volume} {48}} (\bibinfo {year} {2014})}\BibitemShut {NoStop}%
\bibitem [{\citenamefont {Silvi}\ \emph {et~al.}(2019)\citenamefont {Silvi}, \citenamefont {Tschirsich}, \citenamefont {Gerster}, \citenamefont {Jünemann}, \citenamefont {Jaschke}, \citenamefont {Rizzi},\ and\ \citenamefont {Montangero}}]{Silvi2019}%
  \BibitemOpen
  \bibfield  {author} {\bibinfo {author} {\bibfnamefont {P.}~\bibnamefont {Silvi}}, \bibinfo {author} {\bibfnamefont {F.}~\bibnamefont {Tschirsich}}, \bibinfo {author} {\bibfnamefont {M.}~\bibnamefont {Gerster}}, \bibinfo {author} {\bibfnamefont {J.}~\bibnamefont {Jünemann}}, \bibinfo {author} {\bibfnamefont {D.}~\bibnamefont {Jaschke}}, \bibinfo {author} {\bibfnamefont {M.}~\bibnamefont {Rizzi}}, \ and\ \bibinfo {author} {\bibfnamefont {S.}~\bibnamefont {Montangero}},\ }\bibfield  {title} {\enquote {\bibinfo {title} {The tensor networks anthology: Simulation techniques for many-body quantum lattice systems},}\ }\href {https://doi.org/10.21468/SciPostPhysLectNotes.8} {\bibfield  {journal} {\bibinfo  {journal} {SciPost Phys. Lect. Notes}\ }\textbf {\bibinfo {volume} {8}} (\bibinfo {year} {2019})}\BibitemShut {NoStop}%
\bibitem [{\citenamefont {Hill}\ and\ \citenamefont {Wootters}(1997)}]{Concurrence1}%
  \BibitemOpen
  \bibfield  {author} {\bibinfo {author} {\bibfnamefont {S.~A.}\ \bibnamefont {Hill}}\ and\ \bibinfo {author} {\bibfnamefont {W.~K.}\ \bibnamefont {Wootters}},\ }\bibfield  {title} {\enquote {\bibinfo {title} {Entanglement of a pair of quantum bits},}\ }\href {https://doi.org/10.1103/PhysRevLett.78.5022} {\bibfield  {journal} {\bibinfo  {journal} {Phys. Rev. Lett.}\ }\textbf {\bibinfo {volume} {78}} (\bibinfo {year} {1997})}\BibitemShut {NoStop}%
\bibitem [{\citenamefont {Wootters}(1998)}]{Concurrence2}%
  \BibitemOpen
  \bibfield  {author} {\bibinfo {author} {\bibfnamefont {W.~K.}\ \bibnamefont {Wootters}},\ }\bibfield  {title} {\enquote {\bibinfo {title} {Entanglement of formation of an arbitrary state of two qubits},}\ }\href {https://doi.org/10.1103/PhysRevLett.80.2245} {\bibfield  {journal} {\bibinfo  {journal} {Phys. Rev. Lett.}\ }\textbf {\bibinfo {volume} {80}} (\bibinfo {year} {1998})}\BibitemShut {NoStop}%
\bibitem [{\citenamefont {Terhal}\ and\ \citenamefont {Vollbrecht}(2000)}]{Terhal2000}%
  \BibitemOpen
  \bibfield  {author} {\bibinfo {author} {\bibfnamefont {Barbara~M.}\ \bibnamefont {Terhal}}\ and\ \bibinfo {author} {\bibfnamefont {Karl Gerd~H.}\ \bibnamefont {Vollbrecht}},\ }\bibfield  {title} {\enquote {\bibinfo {title} {Entanglement of formation for isotropic states},}\ }\href {https://doi.org/10.1103/PhysRevLett.85.2625} {\bibfield  {journal} {\bibinfo  {journal} {Phys. Rev. Lett.}\ }\textbf {\bibinfo {volume} {85}} (\bibinfo {year} {2000})}\BibitemShut {NoStop}%
\bibitem [{\citenamefont {Vollbrecht}\ and\ \citenamefont {Werner}(2001)}]{Vollbrecht2001}%
  \BibitemOpen
  \bibfield  {author} {\bibinfo {author} {\bibfnamefont {K.~G.~H.}\ \bibnamefont {Vollbrecht}}\ and\ \bibinfo {author} {\bibfnamefont {R.~F.}\ \bibnamefont {Werner}},\ }\bibfield  {title} {\enquote {\bibinfo {title} {Entanglement measures under symmetry},}\ }\href {https://doi.org/10.1103/PhysRevA.64.062307} {\bibfield  {journal} {\bibinfo  {journal} {Phys. Rev. A}\ }\textbf {\bibinfo {volume} {64}} (\bibinfo {year} {2001})}\BibitemShut {NoStop}%
\bibitem [{\citenamefont {Verstraete}\ \emph {et~al.}(2004)\citenamefont {Verstraete}, \citenamefont {García-Ripoll},\ and\ \citenamefont {Cirac}}]{Verstraete2004}%
  \BibitemOpen
  \bibfield  {author} {\bibinfo {author} {\bibfnamefont {F.}~\bibnamefont {Verstraete}}, \bibinfo {author} {\bibfnamefont {J.~J.}\ \bibnamefont {García-Ripoll}}, \ and\ \bibinfo {author} {\bibfnamefont {J.~I.}\ \bibnamefont {Cirac}},\ }\bibfield  {title} {\enquote {\bibinfo {title} {Matrix product density operators: Simulation of finite-temperature and dissipative systems},}\ }\href {https://doi.org/10.1103/PhysRevLett.93.207204} {\bibfield  {journal} {\bibinfo  {journal} {Phys. Rev. Lett.}\ }\textbf {\bibinfo {volume} {93}} (\bibinfo {year} {2004})}\BibitemShut {NoStop}%
\bibitem [{\citenamefont {Zwolak}\ and\ \citenamefont {Vidal}(2004)}]{Zwolak2004}%
  \BibitemOpen
  \bibfield  {author} {\bibinfo {author} {\bibfnamefont {Michael}\ \bibnamefont {Zwolak}}\ and\ \bibinfo {author} {\bibfnamefont {Guifr\'e}\ \bibnamefont {Vidal}},\ }\bibfield  {title} {\enquote {\bibinfo {title} {Mixed-state dynamics in one-dimensional quantum lattice systems: A time-dependent superoperator renormalization algorithm},}\ }\href {\doibase 10.1103/PhysRevLett.93.207205} {\bibfield  {journal} {\bibinfo  {journal} {Phys. Rev. Lett.}\ }\textbf {\bibinfo {volume} {93}},\ \bibinfo {pages} {207205} (\bibinfo {year} {2004})}\BibitemShut {NoStop}%
\bibitem [{\citenamefont {Weiss}\ \emph {et~al.}(2018)\citenamefont {Weiss}, \citenamefont {Gerster}, \citenamefont {Jaschke}, \citenamefont {Silvi},\ and\ \citenamefont {Montangero}}]{Weiss2018}%
  \BibitemOpen
  \bibfield  {author} {\bibinfo {author} {\bibfnamefont {Werner}\ \bibnamefont {Weiss}}, \bibinfo {author} {\bibfnamefont {Matthias}\ \bibnamefont {Gerster}}, \bibinfo {author} {\bibfnamefont {Daniel}\ \bibnamefont {Jaschke}}, \bibinfo {author} {\bibfnamefont {Pietro}\ \bibnamefont {Silvi}}, \ and\ \bibinfo {author} {\bibfnamefont {Simone}\ \bibnamefont {Montangero}},\ }\bibfield  {title} {\enquote {\bibinfo {title} {Kibble-zurek scaling of the one-dimensional bose-hubbard model at finite temperatures},}\ }\href {https://doi.org/10.1103/PhysRevA.98.063601} {\bibfield  {journal} {\bibinfo  {journal} {Phys. Rev. A}\ }\textbf {\bibinfo {volume} {98}} (\bibinfo {year} {2018})}\BibitemShut {NoStop}%
\bibitem [{\citenamefont {Haegeman}\ \emph {et~al.}(2011)\citenamefont {Haegeman}, \citenamefont {Cirac}, \citenamefont {Osborne}, \citenamefont {Pižorn}, \citenamefont {Verschelde},\ and\ \citenamefont {Verstraete}}]{Haegeman2011}%
  \BibitemOpen
  \bibfield  {author} {\bibinfo {author} {\bibfnamefont {Jutho}\ \bibnamefont {Haegeman}}, \bibinfo {author} {\bibfnamefont {J.~Ignacio}\ \bibnamefont {Cirac}}, \bibinfo {author} {\bibfnamefont {Tobias~J.}\ \bibnamefont {Osborne}}, \bibinfo {author} {\bibfnamefont {Iztok}\ \bibnamefont {Pižorn}}, \bibinfo {author} {\bibfnamefont {Henri}\ \bibnamefont {Verschelde}}, \ and\ \bibinfo {author} {\bibfnamefont {Frank}\ \bibnamefont {Verstraete}},\ }\bibfield  {title} {\enquote {\bibinfo {title} {Time-dependent variational principle for quantum lattices},}\ }\href {https://doi.org/10.1103/PhysRevLett.107.070601} {\bibfield  {journal} {\bibinfo  {journal} {Phys. Rev. Lett.}\ }\textbf {\bibinfo {volume} {107}} (\bibinfo {year} {2011})}\BibitemShut {NoStop}%
\bibitem [{\citenamefont {Haegeman}\ \emph {et~al.}(2016)\citenamefont {Haegeman}, \citenamefont {Lubich}, \citenamefont {Oseledets}, \citenamefont {Vandereycken},\ and\ \citenamefont {Verstraete}}]{Haegeman2016}%
  \BibitemOpen
  \bibfield  {author} {\bibinfo {author} {\bibfnamefont {Jutho}\ \bibnamefont {Haegeman}}, \bibinfo {author} {\bibfnamefont {Christian}\ \bibnamefont {Lubich}}, \bibinfo {author} {\bibfnamefont {Ivan}\ \bibnamefont {Oseledets}}, \bibinfo {author} {\bibfnamefont {Bart}\ \bibnamefont {Vandereycken}}, \ and\ \bibinfo {author} {\bibfnamefont {Frank}\ \bibnamefont {Verstraete}},\ }\bibfield  {title} {\enquote {\bibinfo {title} {Unifying time evolution and optimization with matrix product states},}\ }\href {https://doi.org/10.1103/PhysRevB.94.165116} {\bibfield  {journal} {\bibinfo  {journal} {Phys. Rev. B}\ }\textbf {\bibinfo {volume} {94}} (\bibinfo {year} {2016})}\BibitemShut {NoStop}%
\bibitem [{\citenamefont {Paeckel}\ \emph {et~al.}(2019)\citenamefont {Paeckel}, \citenamefont {Köhler}, \citenamefont {Swoboda}, \citenamefont {Manmana}, \citenamefont {Schollwöck},\ and\ \citenamefont {Hubig}}]{Paeckel2019}%
  \BibitemOpen
  \bibfield  {author} {\bibinfo {author} {\bibfnamefont {Sebastian}\ \bibnamefont {Paeckel}}, \bibinfo {author} {\bibfnamefont {Thomas}\ \bibnamefont {Köhler}}, \bibinfo {author} {\bibfnamefont {Andreas}\ \bibnamefont {Swoboda}}, \bibinfo {author} {\bibfnamefont {R.~Salvatore}\ \bibnamefont {Manmana}}, \bibinfo {author} {\bibfnamefont {Ulrich}\ \bibnamefont {Schollwöck}}, \ and\ \bibinfo {author} {\bibfnamefont {Claudius}\ \bibnamefont {Hubig}},\ }\bibfield  {title} {\enquote {\bibinfo {title} {Time-evolution methods for matrix-product states},}\ }\href {https://doi.org/10.1016/j.aop.2019.167998} {\bibfield  {journal} {\bibinfo  {journal} {Annals of Physics}\ }\textbf {\bibinfo {volume} {411}} (\bibinfo {year} {2019})}\BibitemShut {NoStop}%
\bibitem [{\citenamefont {Jaschke}\ \emph {et~al.}(2018)\citenamefont {Jaschke}, \citenamefont {Montangero},\ and\ \citenamefont {Carr}}]{Jaschke2018}%
  \BibitemOpen
  \bibfield  {author} {\bibinfo {author} {\bibfnamefont {Daniel}\ \bibnamefont {Jaschke}}, \bibinfo {author} {\bibfnamefont {Simone}\ \bibnamefont {Montangero}}, \ and\ \bibinfo {author} {\bibfnamefont {D.~Lincoln}\ \bibnamefont {Carr}},\ }\bibfield  {title} {\enquote {\bibinfo {title} {One-dimensional many-body entangled open quantum systems with tensor network methods},}\ }\href {https://doi.org/10.1088/2058-9565/aae724} {\bibfield  {journal} {\bibinfo  {journal} {Quantum Science and Technology}\ }\textbf {\bibinfo {volume} {4}} (\bibinfo {year} {2018})}\BibitemShut {NoStop}%
\bibitem [{\citenamefont {Weimer}\ and\ \citenamefont {Büchler}(2010)}]{Weimer2010}%
  \BibitemOpen
  \bibfield  {author} {\bibinfo {author} {\bibfnamefont {Hendrik}\ \bibnamefont {Weimer}}\ and\ \bibinfo {author} {\bibfnamefont {Hans~Peter}\ \bibnamefont {Büchler}},\ }\bibfield  {title} {\enquote {\bibinfo {title} {Two-stage melting in systems of strongly interacting {Rydberg} atoms},}\ }\href {https://doi.org/10.1103/PhysRevLett.105.230403} {\bibfield  {journal} {\bibinfo  {journal} {Phys. Rev. Lett.}\ }\textbf {\bibinfo {volume} {105}} (\bibinfo {year} {2010})}\BibitemShut {NoStop}%
\bibitem [{\citenamefont {Samajdar}\ \emph {et~al.}(2018)\citenamefont {Samajdar}, \citenamefont {Choi}, \citenamefont {Pichler}, \citenamefont {Lukin},\ and\ \citenamefont {Sachdev}}]{Samajdar2018}%
  \BibitemOpen
  \bibfield  {author} {\bibinfo {author} {\bibfnamefont {Rhine}\ \bibnamefont {Samajdar}}, \bibinfo {author} {\bibfnamefont {Soonwon}\ \bibnamefont {Choi}}, \bibinfo {author} {\bibfnamefont {Hannes}\ \bibnamefont {Pichler}}, \bibinfo {author} {\bibfnamefont {Mikhail~D.}\ \bibnamefont {Lukin}}, \ and\ \bibinfo {author} {\bibfnamefont {Subir}\ \bibnamefont {Sachdev}},\ }\bibfield  {title} {\enquote {\bibinfo {title} {Numerical study of the chiral $\mathds{Z}_3$ quantum phase transition in one spatial dimension},}\ }\href {https://doi.org/10.1103/PhysRevA.98.023614} {\bibfield  {journal} {\bibinfo  {journal} {Phys. Rev. A}\ }\textbf {\bibinfo {volume} {98}} (\bibinfo {year} {2018})}\BibitemShut {NoStop}%
\bibitem [{\citenamefont {Rader}\ and\ \citenamefont {Lauchli}(2019)}]{Rader2019}%
  \BibitemOpen
  \bibfield  {author} {\bibinfo {author} {\bibfnamefont {M.}~\bibnamefont {Rader}}\ and\ \bibinfo {author} {\bibfnamefont {A.~M.}\ \bibnamefont {Lauchli}},\ }\bibfield  {title} {\enquote {\bibinfo {title} {Floating phases in one-dimensional {Rydberg} {I}sing chains},}\ }\href {https://doi.org/10.48550/arXiv.1908.02068} {\bibfield  {journal} {\bibinfo  {journal} {arXiv preprint 1908.02068}\ } (\bibinfo {year} {2019})}\BibitemShut {NoStop}%
\bibitem [{\citenamefont {Yu}\ \emph {et~al.}(2022)\citenamefont {Yu}, \citenamefont {Yang}, \citenamefont {Xu},\ and\ \citenamefont {Xu}}]{Yu2022}%
  \BibitemOpen
  \bibfield  {author} {\bibinfo {author} {\bibfnamefont {X.}~\bibnamefont {Yu}}, \bibinfo {author} {\bibfnamefont {S.}~\bibnamefont {Yang}}, \bibinfo {author} {\bibfnamefont {J.}~\bibnamefont {Xu}}, \ and\ \bibinfo {author} {\bibfnamefont {L.}~\bibnamefont {Xu}},\ }\bibfield  {title} {\enquote {\bibinfo {title} {Fidelity susceptibility as a diagnostic of the commensurate-incommensurate transition: A revisit of the programmable {Rydberg} chain},}\ }\href {https://doi.org/10.1103/PhysRevB.106.165124} {\bibfield  {journal} {\bibinfo  {journal} {Phys. Rev. B}\ }\textbf {\bibinfo {volume} {106}} (\bibinfo {year} {2022})}\BibitemShut {NoStop}%
\bibitem [{\citenamefont {Maceira}\ \emph {et~al.}(2022)\citenamefont {Maceira}, \citenamefont {Chepiga},\ and\ \citenamefont {Mila}}]{Maceira2022}%
  \BibitemOpen
  \bibfield  {author} {\bibinfo {author} {\bibfnamefont {Ivo~A.}\ \bibnamefont {Maceira}}, \bibinfo {author} {\bibfnamefont {Natalia}\ \bibnamefont {Chepiga}}, \ and\ \bibinfo {author} {\bibfnamefont {Frédéric}\ \bibnamefont {Mila}},\ }\bibfield  {title} {\enquote {\bibinfo {title} {Conformal and chiral phase transitions in {Rydberg} chains},}\ }\href {https://doi.org/10.1103/PhysRevResearch.4.043102} {\bibfield  {journal} {\bibinfo  {journal} {Phys. Rev. Research}\ }\textbf {\bibinfo {volume} {4}} (\bibinfo {year} {2022})}\BibitemShut {NoStop}%
\bibitem [{\citenamefont {H\"{o}lzl}\ \emph {et~al.}(2023)\citenamefont {H\"{o}lzl}, \citenamefont {G\"{o}tzelmann}, \citenamefont {Wirth}, \citenamefont {Safronova}, \citenamefont {Weber},\ and\ \citenamefont {Meinert}}]{Holzl2023}%
  \BibitemOpen
  \bibfield  {author} {\bibinfo {author} {\bibfnamefont {C.}~\bibnamefont {H\"{o}lzl}}, \bibinfo {author} {\bibfnamefont {A.}~\bibnamefont {G\"{o}tzelmann}}, \bibinfo {author} {\bibfnamefont {M.}~\bibnamefont {Wirth}}, \bibinfo {author} {\bibfnamefont {M.~S.}\ \bibnamefont {Safronova}}, \bibinfo {author} {\bibfnamefont {S.}~\bibnamefont {Weber}}, \ and\ \bibinfo {author} {\bibfnamefont {F.}~\bibnamefont {Meinert}},\ }\bibfield  {title} {\enquote {\bibinfo {title} {Motional ground-state cooling of single atoms in state-dependent optical tweezers},}\ }\href {https://doi.org/10.1103/PhysRevResearch.5.033093} {\bibfield  {journal} {\bibinfo  {journal} {Phys. Rev. Research}\ }\textbf {\bibinfo {volume} {5}} (\bibinfo {year} {2023})}\BibitemShut {NoStop}%
\bibitem [{\citenamefont {Cataldi}\ \emph {et~al.}(2021)\citenamefont {Cataldi}, \citenamefont {Abedi}, \citenamefont {Magnifico}, \citenamefont {Notarnicola}, \citenamefont {Dalla~Pozza}, \citenamefont {Giovannetti},\ and\ \citenamefont {Montangero}}]{Cataldi2021}%
  \BibitemOpen
  \bibfield  {author} {\bibinfo {author} {\bibfnamefont {Giovanni}\ \bibnamefont {Cataldi}}, \bibinfo {author} {\bibfnamefont {Ashkan}\ \bibnamefont {Abedi}}, \bibinfo {author} {\bibfnamefont {Giuseppe}\ \bibnamefont {Magnifico}}, \bibinfo {author} {\bibfnamefont {Simone}\ \bibnamefont {Notarnicola}}, \bibinfo {author} {\bibfnamefont {Nicola}\ \bibnamefont {Dalla~Pozza}}, \bibinfo {author} {\bibfnamefont {Vittorio}\ \bibnamefont {Giovannetti}}, \ and\ \bibinfo {author} {\bibfnamefont {Simone}\ \bibnamefont {Montangero}},\ }\bibfield  {title} {\enquote {\bibinfo {title} {Hilbert curve vs hilbert space: exploiting fractal 2d covering to increase tensor network efficiency},}\ }\href {https://doi.org/10.22331/q-2021-09-29-556} {\bibfield  {journal} {\bibinfo  {journal} {Quantum}\ }\textbf {\bibinfo {volume} {5}},\ \bibinfo {pages} {556} (\bibinfo {year} {2021})}\BibitemShut {NoStop}%
\bibitem [{\citenamefont {Pirvu}\ \emph {et~al.}(2010)\citenamefont {Pirvu}, \citenamefont {Murg}, \citenamefont {Cirac},\ and\ \citenamefont {Verstraete}}]{Pirvu2010}%
  \BibitemOpen
  \bibfield  {author} {\bibinfo {author} {\bibfnamefont {B.}~\bibnamefont {Pirvu}}, \bibinfo {author} {\bibfnamefont {Valentin}\ \bibnamefont {Murg}}, \bibinfo {author} {\bibfnamefont {J~Ignacio}\ \bibnamefont {Cirac}}, \ and\ \bibinfo {author} {\bibfnamefont {Frank}\ \bibnamefont {Verstraete}},\ }\bibfield  {title} {\enquote {\bibinfo {title} {Matrix product operator representations},}\ }\href {https://doi.org/10.1088/1367-2630/12/2/025012} {\bibfield  {journal} {\bibinfo  {journal} {New Journal of Physics}\ }\textbf {\bibinfo {volume} {12}},\ \bibinfo {pages} {025012} (\bibinfo {year} {2010})}\BibitemShut {NoStop}%
\bibitem [{\citenamefont {Alhambra}\ and\ \citenamefont {Cirac}(2021)}]{Alhambra2021}%
  \BibitemOpen
  \bibfield  {author} {\bibinfo {author} {\bibfnamefont {Álvaro~M.}\ \bibnamefont {Alhambra}}\ and\ \bibinfo {author} {\bibfnamefont {J.~Ignacio}\ \bibnamefont {Cirac}},\ }\bibfield  {title} {\enquote {\bibinfo {title} {Locally accurate tensor networks for thermal states and time evolution},}\ }\href {https://doi.org/10.1103/PRXQuantum.2.040331} {\bibfield  {journal} {\bibinfo  {journal} {PRX Quantum}\ }\textbf {\bibinfo {volume} {2}},\ \bibinfo {pages} {040331} (\bibinfo {year} {2021})}\BibitemShut {NoStop}%
\bibitem [{\citenamefont {Ballarin}\ \emph {et~al.}(2024)\citenamefont {Ballarin}, \citenamefont {Cataldi}, \citenamefont {Costantini}, \citenamefont {Jaschke}, \citenamefont {Magnifico}, \citenamefont {Montangero}, \citenamefont {Notarnicola}, \citenamefont {Pagano}, \citenamefont {Pavešić}, \citenamefont {Rigobello}, \citenamefont {Reinić}, \citenamefont {Scarlatella},\ and\ \citenamefont {Silvi}}]{QuantumTea2024}%
  \BibitemOpen
  \bibfield  {author} {\bibinfo {author} {\bibfnamefont {Marco}\ \bibnamefont {Ballarin}}, \bibinfo {author} {\bibfnamefont {Giovanni}\ \bibnamefont {Cataldi}}, \bibinfo {author} {\bibfnamefont {Aurora}\ \bibnamefont {Costantini}}, \bibinfo {author} {\bibfnamefont {Daniel}\ \bibnamefont {Jaschke}}, \bibinfo {author} {\bibfnamefont {Giuseppe}\ \bibnamefont {Magnifico}}, \bibinfo {author} {\bibfnamefont {Simone}\ \bibnamefont {Montangero}}, \bibinfo {author} {\bibfnamefont {Simone}\ \bibnamefont {Notarnicola}}, \bibinfo {author} {\bibfnamefont {Alice}\ \bibnamefont {Pagano}}, \bibinfo {author} {\bibfnamefont {Luka}\ \bibnamefont {Pavešić}}, \bibinfo {author} {\bibfnamefont {Marco}\ \bibnamefont {Rigobello}}, \bibinfo {author} {\bibfnamefont {Nora}\ \bibnamefont {Reinić}}, \bibinfo {author} {\bibfnamefont {Simone}\ \bibnamefont {Scarlatella}}, \ and\ \bibinfo {author} {\bibfnamefont {Pietro}\ \bibnamefont {Silvi}},\ }\href {https://doi.org/10.5281/zenodo.10498929} {\enquote {\bibinfo {title} {Quantum {TEA}:
  qtealeaves},}\ } (\bibinfo {year} {2024})\BibitemShut {NoStop}%
\bibitem [{\citenamefont {Reinić}\ \emph {et~al.}(2024{\natexlab{a}})\citenamefont {Reinić}, \citenamefont {Jaschke}, \citenamefont {Wanisch}, \citenamefont {Silvi},\ and\ \citenamefont {Montangero}}]{Zenodo}%
  \BibitemOpen
  \bibfield  {author} {\bibinfo {author} {\bibfnamefont {Nora}\ \bibnamefont {Reinić}}, \bibinfo {author} {\bibfnamefont {Daniel}\ \bibnamefont {Jaschke}}, \bibinfo {author} {\bibfnamefont {Darvin}\ \bibnamefont {Wanisch}}, \bibinfo {author} {\bibfnamefont {Pietro}\ \bibnamefont {Silvi}}, \ and\ \bibinfo {author} {\bibfnamefont {Simone}\ \bibnamefont {Montangero}},\ }\href {https://doi.org/10.5281/zenodo.11349566} {\enquote {\bibinfo {title} {Datasets and simulation scripts for {"Finite-temperature {R}ydberg arrays: quantum phases and entanglement characterization"}},}\ } (\bibinfo {year} {2024}{\natexlab{a}})\BibitemShut {NoStop}%
\bibitem [{\citenamefont {Reinić}\ \emph {et~al.}(2024{\natexlab{b}})\citenamefont {Reinić}, \citenamefont {Jaschke}, \citenamefont {Wanisch}, \citenamefont {Silvi},\ and\ \citenamefont {Montangero}}]{Figshare}%
  \BibitemOpen
  \bibfield  {author} {\bibinfo {author} {\bibfnamefont {Nora}\ \bibnamefont {Reinić}}, \bibinfo {author} {\bibfnamefont {Daniel}\ \bibnamefont {Jaschke}}, \bibinfo {author} {\bibfnamefont {Darvin}\ \bibnamefont {Wanisch}}, \bibinfo {author} {\bibfnamefont {Pietro}\ \bibnamefont {Silvi}}, \ and\ \bibinfo {author} {\bibfnamefont {Simone}\ \bibnamefont {Montangero}},\ }\href {https://doi.org/10.6084/m9.figshare.25909111} {\enquote {\bibinfo {title} {Figures for "{Finite-temperature Rydberg arrays: quantum phases and entanglement characterization}"},}\ } (\bibinfo {year} {2024}{\natexlab{b}})\BibitemShut {NoStop}%
\bibitem [{\citenamefont {Mezzadri}(2007)}]{Mezzadri2007}%
  \BibitemOpen
  \bibfield  {author} {\bibinfo {author} {\bibfnamefont {Francesco}\ \bibnamefont {Mezzadri}},\ }\bibfield  {title} {\enquote {\bibinfo {title} {How to generate random matrices from the classical compact groups},}\ }\href {https://www.ams.org/notices/200705/fea-mezzadri-web.pdf} {\bibfield  {journal} {\bibinfo  {journal} {Notices of the American Mathematical Society}\ }\textbf {\bibinfo {volume} {54}},\ \bibinfo {pages} {592--604} (\bibinfo {year} {2007})}\BibitemShut {NoStop}%
\end{thebibliography}%

\end{document}